\documentstyle[latexsym,amssymb,aps,floats,eqsecnum,
preprint,amsfonts]{revtex}
\def\laq{\raise 0.4ex\hbox{$<$}\kern -0.8em\lower 0.62 ex\hbox{$\sim$}}
\def\gaq{\raise 0.4ex\hbox{$>$}\kern -0.7em\lower 0.62 ex\hbox{$\sim$}}
\input epsf
\begin{document}

\bibliographystyle{unsrt}

\title{Zero modes of six-dimensional Abelian vortices}

\author{Massimo Giovannini\footnote{Electronic address: 
massimo.giovannini@ipt.unil.ch}, Jean-Vincent Le B\'e and  
St\'ephane Riederer }

\address{{\it Institute of Theoretical Physics, 
University of Lausanne}}
\address{{\it BSP-1015 Dorigny, Lausanne, Switzerland}}

\maketitle
\begin{abstract}
We analyze the fluctuations of 
Nielsen-Olesen vortices arising in 
the six-dimensional Abelian-Higgs model. The regular  
geometry generated by the defect breaks spontaneously 
six-dimensional Poincar\'e symmetry leading to a warped 
space-time with finite four-dimensional Planck mass. 
As a consequence, 
the zero mode of the spin two fluctuations of the geometry
is always localized but the graviphoton fields, 
corresponding to spin one metric fluctuations,
give rise to zero modes which are not localized 
either because of their behaviour at infinity or because of their 
behaviour near the core of the vortex. A similar situation 
occurs for spin zero fluctuations.
Gauge field fluctuations exhibit a localized zero mode.
\end{abstract}
\vskip 1truecm
\centerline{\bf To appear in Classical and Quantum Gravity}
\newpage

\renewcommand{\theequation}{1.\arabic{equation}}
\setcounter{equation}{0}
\section{Introduction}
In the absence of gravitational interactions, 
five-dimensional domain-wall solutions allow 
the localization of fermionic zero modes \cite{m1}. 
The increase of the dimensionality 
of space-time represents a tool in order to 
obtain an effective lower dimensional theory 
where chiral fermions may be successfully localized. 
The  chiral fermionic 
zero modes is still present if the five-dimensional 
continuous space is replaced by a lattice \cite{kap}.
Chiral symmetry can then  be realized in different ways avoiding 
the known problem of doubling of fermionic degrees of freedom \cite{nin}.
A crucial ingredient, in this context, was the use of the so-called
Wilson-Ginzparg relation \cite{wg}. 
If we move from five to six dimensions, chiral fermions 
can still be localized \cite{tro1,tro2} and the structure 
of the zero modes gets more realistic. 

If gravitational interactions are consistently included \cite{m2,ak,vis,ran},
infinite  extra-dimensions do not only lead to the localization of chiral 
fermions but also to the localization of gravity itself \cite{rs1,rs2}.
Gravitons can be localized in  five dimensional 
${\rm AdS}$ space-times with a brane source \cite{rs1,rs2} and 
explicit physical (thick) brane solutions, compatible 
with ${\rm AdS}_{5}$ geometries,  have  been derived
using  scalar domain-walls 
\cite{kt1,kt2,gremm1,gremm2,free,free2} breaking
five dimensional Poincar\'e invariance but leaving unbroken 
Poincar\'e group in four space-time dimensions. 
Fields of various spin, coming from the 
fluctuations of the geometry itself, can be classified, according 
to four-dimensional Poincar\'e transformations, into scalars, vectors 
and tensors. In the case of scalar domain-walls 
neither the scalar nor the vector 
modes of the geometry are localized \cite{mg1,mg2} if the 
four-dimensional Planck mass is finite. A field of given spin 
is localized if its
the corresponding zero modes are normalizable as a function of 
the bulk coordinates describing the geometry in the internal space.

By going from five to six dimensions not only the volume 
of the internal space gets larger but also supplementary
metric fluctuations arise. 
Thin strings \cite{def}--\cite{gs},  monopoles
\cite{grs}--\cite{Dvali:2000ty} or  instantons 
\cite{Randjbar-Daemi:1983qa}--\cite{def3a}  
( in  six, seven or eight dimensional space-times) 
have been studied mainly looking at the 
localization properties of the tensor modes of the geometry.

In six-dimensions, physical brane solutions (including  
background gauge fields) have been recently analyzed \cite{mhm1} in the 
context of the Abelian-Higgs model and later 
generalized to the case when higher derivative terms are present in the 
gravity part of the action \cite{mhm2}.  
These solutions 
are the analog of the Nielsen-Olesen vortices \cite{no} 
in a six-dimensional 
warped geometry where the two transverse dimensions play the 
r\^ole of the radial and angular coordinates 
defining the location of the Abelian string. 

The purpose of the present investigation is the analysis of the zero modes 
of the vortices arising in the six-dimensional Abelian-Higgs model.  
While the spin two of the geometry 
are decoupled from the very beginning, the spin one fluctuations 
of the metric are coupled with the vector fluctuations arising in the gauge 
sector. Finally the scalar fluctuations of the metric are coupled with 
the spin zero fluctuations arising both  from the Higgs sector and  
gauge sectors. 

Following the formalism developed in \cite{mg1,mg2}, the 
invariance for infinitesimal coordinate transformations, can be 
used in order to address the problem in a coordinate independent 
way. This is the spirit of Bardeen formalism which 
was originally formulated in four-dimensions \cite{bar}, later generalized to 
five-dimensional warped geometries \cite{mg1,mg2} and now 
applied in order to discuss the zero modes of six-dimensional Abelian 
vortices.

The plan of our paper is then the following. In Section II the main 
features of the six-dimensional Abelian vortices will be summarized. 
Particular attention will be given  to those properties of the background 
entering directly the discussion of the zero modes. In Section III the 
gauge-invariant fluctuations of the six-dimensional geometry will be 
introduced and classified. The general system of the fluctuations 
will be presented in Section IV. 
The tensor zero modes will be analyzed in Section V, while 
Section VI and VII will deal, respectively, with the gauge and scalar sectors. 
Section VIII contains our concluding remarks. In the Appendices
various technical results have been summarized.

\renewcommand{\theequation}{2.\arabic{equation}}
\setcounter{equation}{0}
\section{Abelian vortices in six dimensions}

The gravitating 
Abelian-Higgs model with cosmological constant in the bulk \footnote{ Notice 
that ${\cal D}_{A}=\nabla_{A}-ieA_{A}$ is the gauge covariant derivative, 
while $\nabla_{A}$ is the generally covariant derivative. Latin (uppercase) 
indices run over the six dimensional space-time. Greek indices run over 
the four (Poincar\'e invariant) dimensions. Latin (lowercase) indices run 
over the two-dimensional transverse space. }
\begin{equation}
S=\int
d^6x\sqrt{-G}\biggl[ - \frac{ R}{2 \kappa} - \Lambda+
\frac{1}{2}G^{A B}({\cal D}_{A}\varphi)^*{\cal D}_{B}\varphi-\frac{1}{4}
F_{AB}F^{AB}
-\frac{\lambda}{4}\left(\varphi^*\varphi-v^2\right)^2\biggr]~,
\label{a1}
\end{equation}
will be studied 
in a six-dimensional warped geometry whose line element can be written as 
\begin{equation}
ds^2=G_{AB} dx^{A} dx^{B} =
M^2(\rho)\eta_{\mu\nu}dx^\mu dx^\nu-d\rho^2-L(\rho)^2d\theta^2,
\label{metric}
\end{equation}
where $\rho$ is the bulk radius, $\eta_{\mu\nu}$ is the
four-dimensional Minkowski metric and $M(\rho)$, $L(\rho)$ are the warp
factors. In Eqs. (\ref{a1}) $v$ is the vacuum expectation 
value of the Higgs field, $\lambda$ is the self-coupling constant and $e$ 
is the gauge coupling. The metric of Eq. (\ref{metric}) breaks naturally 
Poincar\'e symmetry in six dimensions.

From the action of Eq. (\ref{a1}) the corresponding 
equations of motion can be derived and they are 
 \begin{eqnarray}
&&G^{A B}\nabla_{A}\nabla_{B}\varphi
-e^2A_{A}A^{A} \varphi -ieA_{A}\partial^{A}\varphi-ie\nabla_{A}(A^{A}\varphi)
+\lambda (\varphi^*\varphi-v^2)\varphi=0,
\label{ph}\\
&&\nabla_{A} F^{AB}=-e^2A^{B}\varphi^*\varphi+\frac{ie}{2}
\left(\varphi\partial^{B}\varphi^*-\varphi^*\partial^{B}\varphi\right),
\label{A}\\
&& R_{AB}-\frac{1}{2}G_{AB}R = \kappa\left(T_{AB}+\Lambda G_{AB}\right),
\label{R}
\end{eqnarray}
where 
\begin{eqnarray}
T_{AB}&=&  \frac{1}{2}\left[({\cal D}_A\varphi)^*{\cal D}_{B}\varphi 
+({\cal D}_B\varphi)^*{\cal D}_A\varphi\right]
-\,F_{AC}{F_B}^{C} 
\nonumber\\
&-& G_{A B} \biggl[ \frac{1}{2}({\cal D}_{A}\varphi)^*
{\cal D}^{A}\varphi-\frac{1}{4}
F_{M N}F^{M N}-\frac{\lambda}{4}\left(\varphi^*\varphi-v^2\right)^2\biggr].
\label{enmom}
\end{eqnarray}

The vortex ansatz, characterized by the winding $n$,
\begin{eqnarray} 
&& \phi(\rho,\theta) =vf(\rho)e^{i\,n\,\theta},
\nonumber\\ 
&&A_{\theta}(\rho,\theta)  =\frac{1}{e}[\,n\,-\,P(\rho)] ~,
\label{NO}
\end{eqnarray}
breaks naturally the $U(1)$  symmetry.
Inserting Eqs. (\ref{metric}) and  (\ref{NO}) into 
Eqs. (\ref{ph})--(\ref{R}), the resulting system 
only depends upon the bulk radius.  
It is useful, for practical purposes, 
to define the following set of
dimension-less quantities \footnote{Notice that 
the Higgs boson and vector masses are, in our definitions, 
$m_{H} = \sqrt{2 \lambda} ~ v$ and $ m_{V} = e v$.} 
\begin{equation}
\nu = \kappa v^2, ~~~~~ \alpha = \frac{e^2}{\lambda} ,~~~~~
\mu = \frac{\kappa \Lambda}{\lambda v^2}.
\label{def}
\end{equation}
whose specific numerical value select a given 
solution in the parameter space of the model.
Using the rescalings of Eq. (\ref{def}), it is natural to 
write Eqs. (\ref{ph})--(\ref{R}) in terms of the 
rescaled bulk radius
 $ x = \,m_{H}\,\rho/\sqrt{2} \equiv \sqrt{\lambda} v \rho$, and 
in terms of the derivatives of the logarithms of the warp factors
\begin{equation}
H(x) = \frac{d \ln{M(x)}}{d x},\,\,\,
F(x) = \frac{d \ln{{\cal L}(x)}}{d x},
\end{equation}
where 
\begin{equation}
{\cal L}(x) = \sqrt{\lambda} v L(x).
\label{calL}
\end{equation}
has been defined.

\begin{figure}
\centerline{\epsfxsize = 10 cm  \epsffile{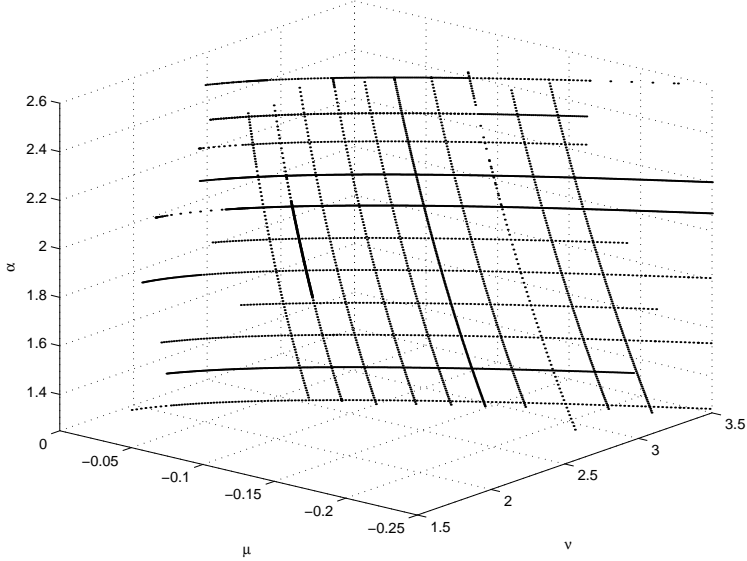}} 
\caption[a]{The parameter space of the vortex solutions.}
\label{F4}
\end{figure}

Eqs. (\ref{ph})--(\ref{R}) become then
\begin{eqnarray} 
&& \frac{ d^2 f}{d x^2} + ( 4 H + F ) \frac{ d f}{d x} +(1 - f^2) f -
\frac{P^2}{{\cal L}^2}f=0,
\label{f1}\\ 
&& \frac{d^2 P}{d x^2}  + ( 4 H - F) \frac{ d P}{ d x} 
  -\alpha f^2 P=0,
\label{p1}\\ 
&& \frac{ d H}{d x} + 3 \frac{ d F}{d x} 
+ F^2 + 6 H^2 + 3 H F  = - \mu - \nu \tau_0,
\label{m1}\\
&& 4 \frac{d H}{ d x}  + 10 H^2 = - \mu - \nu \tau_{\theta},
\label{m2}\\
&& 4 H F + 6 H^2 = -\mu - \nu \tau_{\rho}. 
\label{l1}
\end{eqnarray}
Eqs. (\ref{f1}) and (\ref{p1}) correspond, respectively, 
 to Eqs. (\ref{ph})--(\ref{A}).
The other equations come, respectively, from the 
$(\mu,\nu)$, $(\rho,\rho)$ and $(\theta, \theta)$ components of 
the Einstein equations (\ref{R}). 
The functions  $\tau_{0}$, 
$\tau_{\rho}$ and  $\tau_{\theta}$ denote the 
 components of the energy-momentum tensor
\begin{eqnarray} 
\tau_0(x)  &\equiv& T_{0}^{0}= T_{i}^{i}
= \frac{1}{2} \biggl( \frac{ d f }{ d x}\biggr)^2 
+ \frac{1}{4} ( f^2 -1 )^2 
+ \frac{1}{2 \alpha {\cal L}^2 } \biggl( \frac{d P}{d x}\biggr)^2 
+ \frac{f^2 P^2 }{2 {\cal L}^2},
\label{t0}\\ 
\tau_\rho(x)&\equiv& T_{\rho}^{\rho} = 
-\frac{1}{2} \biggl( \frac{ d f }{ d x}\biggr)^2 
+ \frac{1}{4} ( f^2 -1 )^2 
- \frac{1}{2 \alpha {\cal L}^2 } \biggl( \frac{d P}{d x}\biggr)^2 
+ \frac{f^2 P^2 }{2 {\cal L}^2},
\label{trho}\\ 
\tau_\theta(x)&\equiv& T_{\theta}^{\theta}=
 \frac{1}{2} \biggl( \frac{ d f }{ d x}\biggr)^2 
+ \frac{1}{4} ( f^2 -1 )^2 
- \frac{1}{2 \alpha {\cal L}^2 } \biggl( \frac{d P}{d x}\biggr)^2 
- \frac{f^2 P^2 }{2 {\cal L}^2}.
\label{tth} 
\end{eqnarray}
\begin{figure}
\centerline{\epsfxsize = 10 cm  \epsffile{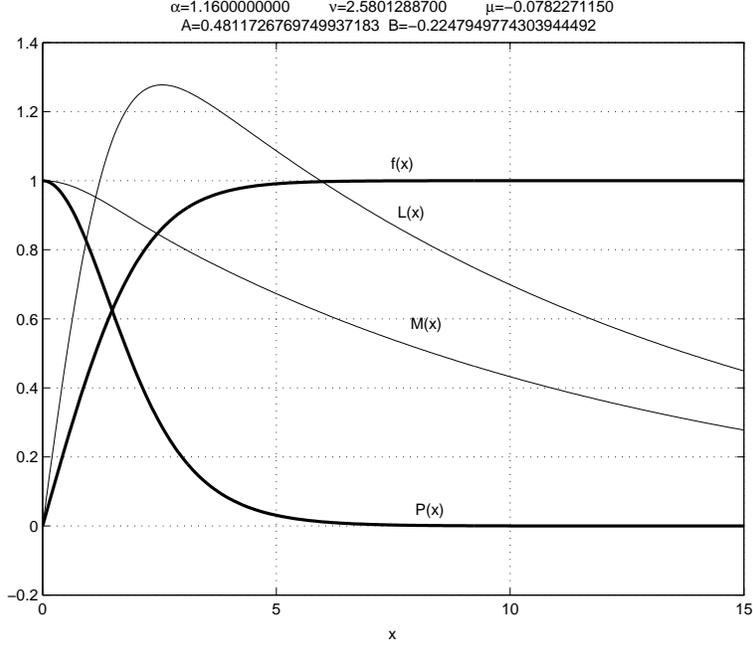}} 
\caption[a]{A typical vortex solution.}
\label{F1}
\end{figure}
Fig. \ref{F1} is representative of a class of 
solutions whose parameter space is reported in Fig. \ref{F4}
in terms of the dimension-less quantities
defined in Eq. (\ref{def}). Each point on the surface of Fig. \ref{F4} 
corresponds to a solution of the type 
of the one reported in Fig. \ref{F1}. The solutions illustrated 
in Fig. \ref{F1} and \ref{F4} have been numerically 
obtained with the techniques described 
in \cite{mhm1,mhm2}.
From Fig. \ref{F1}, it can be appreciated that  the scalar field 
reaches, for large $x$, its vacuum expectation  value and close to
the core of the string the Higgs and gauge fields are  regular:
\begin{eqnarray} 
f(0)=0,&\qquad& \lim_{x\rightarrow \infty} f(x)=1,
\nonumber\\ 
P(0)=n,&\qquad& \lim_{x\rightarrow \infty} P(x)=0. 
\label{boundary}
\end{eqnarray}
The solution of Fig. \ref{F1} corresponds to the 
case of lowest winding , i.e. $n=1$ in Eq. (\ref{NO}), but 
regular solutions with higher winding can be 
easily obtained \cite{mhm1,mhm2}.

\begin{figure}
\centerline{\epsfxsize = 10 cm  \epsffile{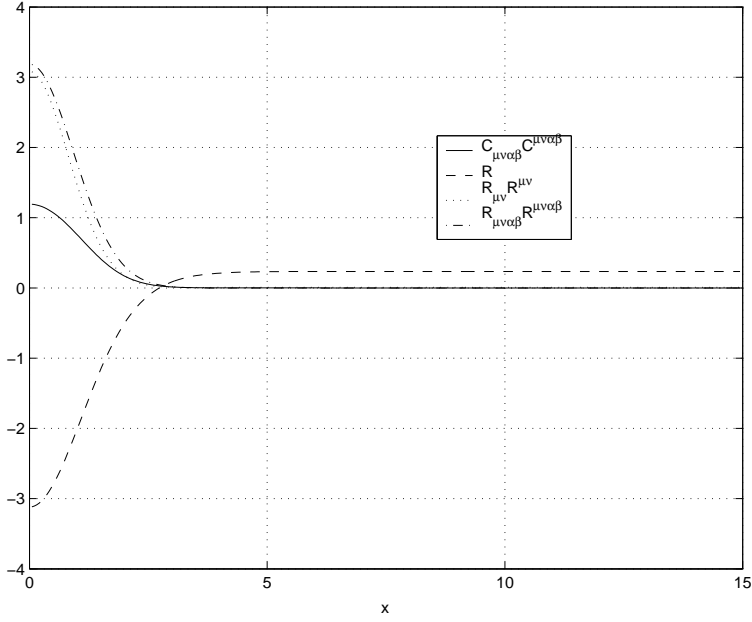}} 
\caption[a]{The curvature invariants computed for the solution 
reported in Fig. (\ref{F1}). }
\label{F2}
\end{figure}

The regularity of the geometry in the core of the string implies
\begin{equation}
\left.\frac{ d M}{d x}\right|_{0} 
= 0~,~~{\cal L}(0) = 0~,~~ \left.\frac{d{\cal L}}{dx}
\right|_{0} = 1~,
\label{boundary2}
\end{equation}
and $M(0) =1$. 
At large distances from the core the 
behaviour of the geometry is ${\rm AdS}_{6}$ space
characterized, in this coordinate system, by exponentially 
decreasing warp factors
\begin{equation}
M(x) \sim e^{ - c x },~~~~{\cal L}(x) \sim e^{- c x} 
\label{ads}
\end{equation}
where $ c = \sqrt{- \mu/10}$.  Since the defects corresponding to the solution
of Fig. \ref{F1} are {\em local}, the corresponding 
energy-momentum tensor goes to zero for large $x$.
Hence, for large $x$, the geometry is determined 
only by the value of the bulk cosmological constant which 
is related to the parameter $\mu$. This feature is also 
illustrated by Fig. \ref{F2} where the components 
of the energy momentum tensors 
are reported in the case of the solution of Fig. \ref{F1}.
\begin{figure}
\centerline{\epsfxsize = 10 cm  \epsffile{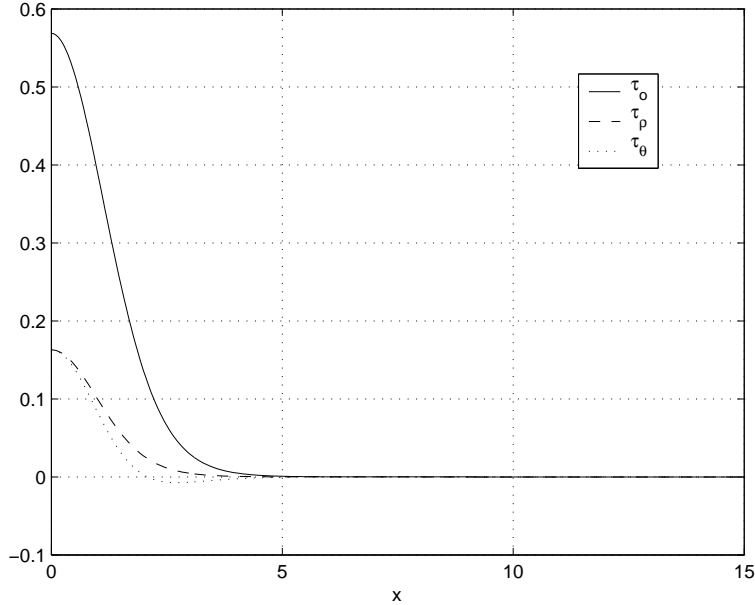}} 
\caption[a]{The components of the energy-momentum tensor reported 
in the case of the solution of Fig. \ref{F1}.}
\label{F3}
\end{figure}

The form of the solutions in the vicinity of the core of the vortex
can be  studied by  expressing the metric functions, together with
the  scalar and gauge fields, as a power series in $x$, i.e. the
dimensionless bulk radius.  Inserting 
the power series
into Eqs.  (\ref{f1})--(\ref{l1}) and requiring that the series obeys,
for $x\rightarrow 0$,  the boundary conditions of Eqs.
(\ref{boundary})  the form of the solutions can be determined as a
function of the  parameters of the model:
\begin{eqnarray}
f(x)&\simeq&Ax+\frac{A}{8}\left(\frac{2\mu}{3}+\frac{\nu}{6}
+\frac{2A^2 \nu}{3}- 1 + 2B + \frac{4B^2 
\nu}{3\alpha}\right)x^3,\\
P(x)&\simeq&1+Bx^2,\\
M(x)&\simeq&1+\left(-\frac{\mu}{8}-\frac{\nu}{32}+
\frac{\nu B^2}{4\alpha}\right)x^2,\\
{\cal L}(x)&\simeq&x+\left[\frac{\mu}{12}+\nu\left(\frac{1}{48}
-\frac{5B^2}{6\alpha}-\frac{A^2}{6}\right)\right]x^3.
\label{x=0}
\end{eqnarray}
In Eq. (\ref{x=0}) $A$ and $B$ are two arbitrary constants which cannot 
be determined by the local analysis of the equations of motion. These
constants are to be found by studying, simultaneously, the 
boundary conditions for $f(x)$ and
$P(x)$ at infinity and in the origin. This analysis can be achieved 
by looking at the behaviour of the string tensions. 

By studying 
the relations among the string tensions \cite{mhm1,mhm2}, the specific value 
of $B$ required in order to have ${\rm AdS}_{6}$ at infinity 
and regular geometry in the origin can be obtained.
Indeed, for all the solutions of the family defined by the fine-tuning surface of 
Fig. \ref{F4} we have that 
\begin{equation}
-\frac{\nu}{\alpha {\cal L}}\left.\frac{d P}{ d x }\right|_0=1\,.
\label{mus2}
\end{equation}
According to Eq. (\ref{x=0}), for $x \rightarrow 0$, 
$ P\sim 1 + B x^2$. Using 
Eq. (\ref{mus2}) the expression for $B$ can be exactly computed
\begin{equation}
B= - \frac{\alpha}{2\nu}\,.
\label{B}
\end{equation}

The solutions of the type of Fig. \ref{F1} 
are regular everywhere, not only in the origin or at infinity. 
For this purpose, the 
behaviour of the curvature invariants [i.e. $R_{ABCD}R^{ABCD}$, 
$C_{ABCD}C^{ABCD}$, $R_{AB}R^{AB}$, $R^2$] 
is reported in Fig. \ref{F2} for the solution of Fig. \ref{F1}.
Far from the core 
the Higgs and gauge fields approach their boundary values in an exponential 
way. Taking into account that, for large $x$, the warp factors 
decrease exponentially, the approach of the gauge and Higgs fields to their boundary 
values can be analytically obtained by inserting
\begin{eqnarray}
&& P(x) = \overline{P} + \delta P(x),~~~ \overline{P}\sim 0
\nonumber\\
&& f(x) = \overline{f} - \delta f(x),~~~ \overline{f} \sim 1
\end{eqnarray}
into  Eqs. (\ref{f1})--(\ref{l1}). The background equations imply that
\begin{eqnarray}
&& \delta P(x) \sim e^{ \sigma_{1} x},\,\,\sigma_1 = 
\frac{3 c}{2} \bigl[ 1 \pm \sqrt{1+  \frac{4 \alpha}{9 c^2}}\bigr],
\label{dP}
\nonumber\\
&& \delta f(x) \sim e^{\sigma_2 x},\,\,\,\, \sigma_2 = 
\frac{5 c}{2} \bigl[ 1 \pm \sqrt{ 1 + \frac{8}{25 c^2}}\bigr].
\label{df}
\end{eqnarray}
If  $4\alpha \gg 9 c^2$ (the limit of small bulk cosmological
constant) the solution  is compatible with the gauge field decreasing
asymptotically as  $\delta P \sim e^{-\sqrt{\alpha}x}$.
If  $ 25 c^2 < 8$ the the perturbed  solution goes as  $\delta f \sim
e^{- \sqrt{2} x}$. 

Consider now  the difference between the $(0,0)$ and $(\theta,\theta)$
components of the  Einstein equations, i.e. (\ref{m1}) and (\ref{m2}):
\begin{equation}
\frac{ d (H - F)}{ d x} 
+ ( F + 4 H) (F - H) = - \nu( \tau_0 - \tau_{\theta} ) 
\label{fminh0}
\end{equation}
Multiplying both sides of this equation by $\sqrt{-G}=M^4 {\cal L}$ and integrating 
from $0$ to $x$, the following relation  
\begin{equation}
( H - F)  = \frac{\nu}{\alpha} 
\frac{ P }{{\cal L}^2}\frac{ d P }{d x} - 
\biggl( 1 + \left.
 \frac{\nu}{\alpha {\cal L}} \frac{d P}{d x }
\right|_0\biggr) \frac{1}{M^4 {\cal L}},
\label{tuneq}
\end{equation}
is obtained.
According to 
Eqs. (\ref{mus2})
and (\ref{B}) the boundary term 
in the core disappears and the resulting equation will be 
\begin{equation}
F = H - \frac{\nu}{\alpha} \frac{P~}{{\cal L}^2} \frac{ d P}{d x}.
\label{fminh}
\end{equation}
Eq. (\ref{fminh})  holds for all the family of solutions 
describing vortex configurations (\ref{NO})  with 
${\rm AdS}_{6}$ behaviour at infinity.

The solutions presented in this Section satisfy everywhere (not only for large 
bulk radius) the corresponding equations of motion obtained 
from the action (\ref{a1}). In this sense, the following 
analysis in not general since it relies  on the specific 
brane action of the Abelian-Higgs model. 
Sometimes (see for instance \cite{ner}) solutions 
of this sort are postulated without assuming a specific 
brane action and by further postulating that the gauge-Higgs 
background is absent.
This procedure is not appropriate in the present case since,
as we shall see, the fluctuations 
of gauge and Higgs fields become sources of the fluctuations
of the geometry.

\renewcommand{\theequation}{3.\arabic{equation}}
\setcounter{equation}{0}
\section{Gauge-invariant fluctuations of Abelian vortices}
The fluctuations of the geometry  and of the gauge-Higgs sources 
around the fixed vortex background will now be discussed.
 In \cite{mg1,mg2}
 gauge-invariant techniques were used in order to address the 
fluctuations of five-dimensional 
domain-wall solutions breaking spontaneously 
five-dimensional Poincar\'e invariance.The main physical difference 
between the scalar domain walls in five dimensions and Abelian vortices in 
six dimensions is given by the presence of a gauge field background whose 
fluctuations mix with the graviphoton fields coming from the geometry.
The invariance of the fluctuations of the six-dimensional metric for an 
infinitesimal coordinate transformation around the  vortex background
guarantees that the obtained fluctuations (and their 
related evolution equations) are independent on the specific coordinate 
system and therefore free of spurious gauge modes \cite{bar}. 

\subsection{Basic considerations}

In analogy with what customarily done in the five-dimensional 
situation \cite{rs2,mg1,mg2}, the six-dimensional 
 line element of Eq. (\ref{metric}) can be written as 
\begin{equation}
ds^2 = M^2(w) [ dt^2 - d\vec{x}^2 ] -  L^2(w) [ dw^2 + d\theta^2] .
\label{newm}
\end{equation}
Eqs. (\ref{metric}) 
and  (\ref{newm}) are connected by the usual  differential relation 
\begin{equation}
d\rho = L(w) d w,
\end{equation}
which can be also written as 
\begin{equation}
 d x = {\cal L}(w) d w,
\label{difr}
\end{equation}
if we recall that, according to Eq. (\ref{calL}), 
$x = \sqrt{\lambda } v \rho $ and $ {\cal L}( w) = \sqrt{\lambda} v L(w)$.

A generic fluctuation $J(x^{\mu},w, \theta)$
 of the six-dimensional Abelian vortex 
 will admit derivatives with respect to the ordinary 
four-dimensional coordinates, but also with respect to the two-dimensional 
transverse space. Hence, the following notation will be employed:
\begin{equation}
J' = \frac{\partial J}{\partial w} ,~~~~
\dot{J} = \frac{\partial J}{\partial \theta}.
\end{equation}
Using Eq. (\ref{newm}) into Eqs. (\ref{ph})--(\ref{R}) together with the 
vortex ansatz
\begin{equation}
\varphi = v~e^{ i n \theta} f(w), ~~~~~A_{\theta}(w) = \frac{ n - P(w)}{e}, 
\label{NOd}
\end{equation}
the equations of the background for the Higgs and gauge field are, 
respectively,
\begin{eqnarray}
&& f '' + 4 {\cal H} f' + {\cal L}^2 f ( 1 - f^2) - P^2 f=0,
\label{hg}\\
&& P'' + ( 4 {\cal H} - 2{\cal F})P' - \alpha f^2 P {\cal L}^2 =0,
\label{Pg}
\end{eqnarray}
while Eq. (\ref{R}) leads to 
\begin{eqnarray}
&& 3 {\cal H}' + {\cal F}' + 6 {\cal H}^2 = - \mu {\cal L}^2
- \nu  \biggl[\frac{ {f'}^2}{2} + \frac{1}{4} ( f^2 -1 )^2 {\cal L}^2
+ \frac{{P'}^2}{2 \alpha {\cal L}^2 } + \frac{f^2 P^2 }{2}\biggr],
\label{e1}\\
&& 6 {\cal H}^2 + 4 {\cal H} {\cal F} =- \mu {\cal L}^2
- \nu  \biggl[-\frac{ {f'}^2}{2} + \frac{1}{4} ( f^2 -1 )^2 {\cal L}^2
- \frac{{P'}^2}{2 \alpha {\cal L}^2 } + \frac{f^2 P^2 }{2}\biggr],
\label{e2}\\
&& 4 {\cal H}' - 4 {\cal H} {\cal F} + 10 {\cal H}^2 = - \mu {\cal L}^2
- \nu  \biggl[\frac{ {f'}^2}{2} + \frac{1}{4} ( f^2 -1 )^2 {\cal L}^2
- \frac{{P'}^2}{2 \alpha {\cal L}^2 } - \frac{f^2 P^2 }{2}\biggr].
\label{e3}
\end{eqnarray}
The quantities 
\begin{equation}
{\cal H} = \frac{\partial \ln{M}}{\partial w} , ~~~~~~
{\cal F} = \frac{\partial \ln{L}}{\partial w} . 
\end{equation}
have been defined in full analogy with  Eqs. (\ref{f1})-(\ref{l1}).

Summing up 
Eqs. (\ref{e2}) and (\ref{e3}) we get
\begin{equation}
4 {\cal H}' + 16 {\cal H}^2 = - 2 \mu {\cal L}^2 - 
\nu \biggl[ \frac{1}{2}(f^2 - 1 )^2 {\cal L}^2 
- \frac{{P'}^2}{\alpha {\cal L}^2}\biggr],
\label{comb}
\end{equation}
Subtracting Eq. (\ref{e3}) from Eq. (\ref{e1}) 
\begin{equation}
 {\cal F}' - {\cal H}' + 4 {\cal H}( {\cal F} - {\cal H}) = 
- \nu \biggl[  \frac{{P'}^2}{\alpha {\cal L}^2 } + f^2 P^2  \biggr],
\label{comb2}
\end{equation}
which is the analog of Eq. (\ref{fminh0}) obtained in the 
$x$-parametrization.
In the limit $w \to \infty$ $P'\to 0$ and $P\to 0$. Hence 
from Eq. (\ref{comb2}),  the limit $P' \to 0$ and $P\to 0$ implies 
${\cal H} \to {\cal F}$, namely the ${\rm AdS}_{6}$ space-time.

Integrating Eq. (\ref{comb2}) with the boundary conditions 
fixed by the vortex configuration and by the relations among the string 
tensions the following relation is obtained
\begin{equation}
{\cal H} - {\cal F} = \frac{\nu}{\alpha {\cal L}^2} P P',
\label{hminf}
\end{equation} 
which is the analog of Eq. (\ref{fminh}) obtained in the 
$x$-parametrization.

Recalling that, for $x\to 0$, ${\cal L}(x) \simeq x$ and using Eq. (\ref{difr}),
the limit $x \to 0$ corresponds, in the $w$ parametrization, to the 
limit $w \to - \infty$. For $w \to -\infty$
\begin{eqnarray}
&& {\cal L}(w) \simeq e^{w} + {\cal O}(e^{3 w}),
\nonumber\\
&& M(w) \simeq 1 +   {\cal O}(e^{2 w}),
\nonumber\\
&& f(w) \simeq A e^{w} + {\cal O} (e^{3 w}), 
\nonumber\\
&& P(w) \simeq 1 +  {\cal O}(e^{2 w}),
\label{w-inf}
\end{eqnarray}
From Eq. (\ref{difr}), the limit $ x \to + \infty$ corresponds to $w \to +\infty$
which implies, according to Eq. (\ref{ads}), 
\begin{eqnarray}
&& M(w) \simeq M_0 \frac{1}{ c ~w},
\nonumber\\
&& {\cal L}(w) \simeq {\cal L}_0 \frac{1}{c ~w}.
\label{w+inf}
\label{madsw}
\end{eqnarray}
As usual, $c = \sqrt{- \mu/10}$ is related to the inverse of the 
${\rm AdS}_{6}$ radius and it is determined by the negative cosmological 
constant dominating the solutions far from the core. 
The asymptotic behaviour of the gauge and Higgs field are, in the same limit,
\begin{eqnarray}
&& f(w ) \simeq ( c w)^{ - \gamma_{f}},
\label{fw}\\
&& P(w) \simeq ( c w)^{- \gamma_{P}} ,
\label{Pw} 
\end{eqnarray}
where $\gamma_{f}>0$ and $\gamma_{P}>0$. According to Eqs. (\ref{df})--(\ref{dP})
$ 4 \alpha \gg 9 c^2$ 
and $ 25 c^2 \ll 8$. Hence, $\gamma_{f} \gg 1$ and $\gamma_{P} \gg 1$:
\begin{equation}
\gamma_{f} = \sqrt{2}/c \equiv \sqrt{ - \frac{20}{\mu}}>1,
 ~~~~\gamma_{P} = \sqrt{\alpha}/c \equiv \sqrt{ - \frac{10 \alpha}{\mu}}> 1,
\end{equation}
implying that $f(w)$ and $P(w)$ reach their boundary values at infinity 
faster than $M(w)$ and ${\cal L}(w)$.

\subsection{Scalar, vector and tensor modes of the geometry}
Since four-dimensional Poincar\'e symmetry is unbroken by the presence 
of the vortex background,
the fluctuation of the metric can be 
decomposed in terms of scalar vector and tensor modes 
\begin{equation}
\delta G_{AB}(x^{\mu}, w,\theta) = \delta G^{(S)}_{AB}(x^{\mu}, w, \theta)
 +\delta 
G^{(V)}_{AB}(x^{\mu}, w,\theta) +\delta G^{(T)}_{AB}(x^{\mu}, w,\theta).
\end{equation} 
with respect to  Poincar\'e transformations along the four 
physical dimensions:
\begin{equation}
\delta G_{A B}=\left(\matrix{2M^2 H_{\mu\nu} 
&L M {\cal G}_{\mu}  & L M {\cal B}_{\mu} \cr
L M {\cal G}_{\mu} & 2 L^2 \xi& L^2 \pi\cr
L M {\cal B}_{\mu} & L^2 \pi & 2 L^2 \phi&\cr}\right),
\label{lorf}
\end{equation}
where 
\begin{eqnarray}
&& H_{\mu\nu} = h_{\mu\nu} +\frac{1}{2} 
(\partial_{\mu} f_{\nu} +\partial_{\nu} f_{\mu}) 
+ \eta_{\mu \nu} \psi
+  \partial_{\mu}\partial_{\nu} E,
\nonumber\\
&& {\cal G}_{\mu} = D_{\mu} + \partial_{\mu}  C,
\nonumber\\
&& {\cal B}_{\mu} = Q_{\mu} + \partial_{\mu} {\cal P},
\end{eqnarray}
with 
\begin{eqnarray}
&& \partial_{\mu}h^{\mu}_{\nu} =0,~~~h_{\mu}^{\mu}=0,
\nonumber\\
&&\partial_{\mu}f^{\mu} =0,~~~\partial_{\mu}D^{\mu} =0,~~~
\partial_{\mu}Q^{\mu} =0.
\end{eqnarray}
The tensor  $h_{\mu\nu}$ has five independent components, while 
 $Q_{\mu}$, $f_{\mu}$ and 
$D_{\mu}$ have, overall nine independent components. 
The scalars $C$, ${\cal P}$, $\psi$, $\phi$, $\xi$,  $E$ and $\pi$ correspond 
to seven scalar degrees of freedom. 

Under infinitesimal coordinates transformations around the 
background geometry fixed by the vortex solution
\begin{equation}
x^{A} \rightarrow \tilde{x}^{A} = x^{A} + \epsilon^{A}, 
\label{shift}
\end{equation}
the twenty one degrees of freedom of the perturbed six-dimensional metric 
do transform as
\begin{equation}
\delta \tilde{G}_{A B} = \delta G_{AB} - \nabla_{A} \epsilon_{B} - \nabla_{B}
\epsilon_{A}, 
\label{liederiv}
\end{equation}
where the the Lie 
the covariant derivatives are computed using the background 
metric (\ref{newm}) and where
\begin{equation} 
\epsilon_{A} = ( M^2 \epsilon_{\mu}, - L^2  \epsilon_{w}, - L^2 
\epsilon_{\theta}).
\end{equation}
The shift along the longitudinal coordinates  has a pure vector part and 
a scalar part \footnote{Notice that in order to write down this 
decomposition, the gauge functions should be regular enough (in order 
to guarantee the existence of $\Box^{-1}$) and, in any case, 
not singular.}
\begin{equation}
\epsilon_{\mu} = \partial_{\mu}\epsilon + \zeta_{\mu}.
\end{equation}
The infinitesimal diffeomorphisms  preserving the {\em scalar} nature 
of the fluctuation involve 
$\epsilon$, $\epsilon_{\theta}$ and $\epsilon_{w}$, whereas
the coordinate transformations preserving the {\em vector} 
nature of the fluctuation 
only involve $\zeta_{\mu}$. Apart from the transverse and traceless
tensors , which  do not change 
for infinitesimal gauge transformations (i.e. 
$\tilde{h}_{\mu\nu} = h_{\mu\nu}$), 
from Eq. (\ref{liederiv})  the {\em vector} transform as 
\begin{eqnarray}
&& \tilde{f}_{\mu} = f_{\mu} - \zeta_{\mu},
\label{fl}\\
&&\tilde{D}_{\mu} = D_{\mu} - \frac{M}{L}\zeta_{\mu}'~,
\label{zeta1}\\
&& \tilde{Q}_{\mu} = Q_{\mu} - \frac{M}{L} \dot{\zeta}_{\mu}~, 
\label{zeta2}
\end{eqnarray}
while  the scalars transform as
\begin{eqnarray}
&&\tilde{E} = E - \epsilon,
\label{El}\\
&&\tilde{\psi} = \psi - {\cal H} \epsilon_{w},
\label{psil}\\
&& \tilde{C} = C - \frac{M}{L}\epsilon' + \frac{L}{M}\epsilon_{w},
\label{Cl}\\
&& \tilde{\xi} = \xi + {\cal F} \epsilon_{w} + \epsilon_{w}',
\label{xil}\\
&& \tilde{{\cal P}} = {\cal P} + \frac{L}{M}\epsilon_{\theta} - 
\frac{M}{L} \dot{\epsilon} ,
\\
&& \tilde{\pi} = \pi + \epsilon_{\theta}' + \dot{\epsilon}_{w}, 
\\
&& \tilde{\phi} = \phi + \dot{\epsilon}_{\theta} + {\cal F} \epsilon_{w}. 
\label{phitr}
\end{eqnarray}
Since there are twenty one independent components of the perturbed metric and 
six gauge functions  we can define fifteen gauge-invariant degrees of freedom. 
These fifteen degrees of freedom are decomposed into four gauge-invariant 
scalars,
\begin{eqnarray}
&&\tilde{\Psi} = \tilde{\psi} + {\cal H} \biggl[\frac{M}{L}
 \tilde{ C} - \frac{M^2}{L^2} \tilde{E}'\biggr],
\label{PSI}\\
&& \tilde{\Xi} = \tilde{\xi} - \frac{1}{L} \biggl[ L \biggl( \frac{M}{L} 
\tilde{C} - \frac{M^2}{L^2} \tilde{E}'\biggl)\biggr]',
\\
&&  \tilde{\Phi} = \tilde{\phi} - \biggl[ \frac{M}{L}\tilde{{\cal P}} - 
\frac{M^2}{L^2} \dot{\tilde{E}}\biggr]^{\cdot} - {\cal F} \biggl[\frac{M}{L}
 \tilde{ C} - \frac{M^2}{L^2} \tilde{E}'\biggr],
\\
&& \tilde{\Pi} = \tilde{\pi} -  \biggl[ \frac{M}{L}\tilde{{\cal P}} - 
\frac{M^2}{L^2} \dot{\tilde{E}}\biggr]'-  \biggl[\frac{M}{L}
 \tilde{ C} - \frac{M^2}{L^2} \tilde{E}'\biggr]^{\cdot},
\label{scgi}
\end{eqnarray}
two gauge-invariant divergence-less vectors (corresponding
to six degrees of freedom)
\begin{eqnarray}
&&\tilde{V}_{\mu} = \tilde{D}_{\mu} - \frac{M}{L} \tilde{f}_{\mu}',
\label{V}\\
&& \tilde{Z}_{\mu} = \tilde{Q}_{\mu} - \frac{M}{L} \dot{\tilde{f}}_{\mu},
\label{Z}
\end{eqnarray}
supplemented by  the divergence-less and trace-less 
tensor degrees of freedom  $h_{\mu\nu}$.

\subsection{Gauge-invariant fluctuations of the sources}

As for the fluctuations of the geometry, the fluctuations 
of the Higgs and gauge fields will also transform for infinitesimal 
diffeomorphisms. 
Following the conventions 
\begin{eqnarray}
&& \varphi( x^{\mu}, w, \theta) = \varphi(w, \theta) + 
\chi( x^{\mu}, w, \theta),
\nonumber\\
&& A_{M}( x^{\mu}, w, \theta) = A_{M}(w) + \delta A_{M} (x^{\mu}, w, \theta),
\end{eqnarray}
we have that
\begin{eqnarray}
&&\delta \tilde{A}_{A} = \delta A_{A} - A^{C} \nabla_{A} \epsilon_{C} 
- \epsilon^{B} \nabla_{B} A_{A} ,
\label{svvar}\\
&& \tilde{\chi} = \chi - \epsilon^{A} \partial_{A} \varphi,
\nonumber\\
&& \tilde{\chi}^* = \chi - \epsilon^{A} \partial_{A} \varphi^* .
\end{eqnarray}
Using the convenient notation 
\begin{eqnarray}
&&\delta A_{A}(x^{\mu}, w, \theta) = \frac{1}{e}a_{A}(x^{\mu}, w, \theta),
\label{ff1}\\
&&\delta \varphi(x^{\mu}, w, \theta) = \chi(x^{\mu}, w, \theta),
 = v~e^{i n \theta} ~ g(x^{\mu}, w, \theta)
\label{ff2}\\
&&\delta \varphi(x^{\mu}, w, \theta)^* = \chi(x^{\mu}, w, \theta)^*
 = v~e^{-i n \theta} ~ g(x^{\mu}, w, \theta)^*,
\label{ff3}
\end{eqnarray}
the gauge variation reads, in explicit terms,
\begin{eqnarray}
&&\tilde{a}_{w} = a_{w} - ( n - P) \dot{\epsilon}_{\theta},
\label{awtr}\\
&& \tilde{a}_{\theta} = a_{\theta} - ( n - P) \dot{\epsilon}_{\theta} 
+ \epsilon_{w} P', 
\nonumber\\
&& \tilde{a} = a - ( n - P) \epsilon_{\theta}, 
\nonumber\\
&& \tilde{{\cal A}}_{\mu} = {\cal A}_{\mu}.
\label{amutr}
\end{eqnarray}
where the vector fluctuation $a_{\mu}$ 
\begin{equation}
a_{\mu} = {\cal A}_{\mu} + \partial_{\mu } a, 
\end{equation}
has been decomposed in a divergence-less part ${\cal A}_{\mu}$ (i.e.  
$\partial_{\mu} {\cal A}^{\mu} =0$), transforming as a 
pure Poincar\'e vector,  and a divergence full part $a$ transforming as 
a scalar.  Thanks to the 
symmetry of the background solutions (whose 
only non-vanishing component is $A_{\theta}$), 
 the pure vector fluctuation, i.e. 
${\cal A}_{\mu}$,  is automatically
invariant for infinitesimal coordinate transformations as implied by 
Eq. (\ref{svvar}).

The other components (i.e. $a$, $a_{w}$ and $a_{\theta}$) do change 
for infinitesimal coordinate transformations. Recalling  the explicit form 
of gauge variation of the metric components given in  
Eqs. (\ref{El})--(\ref{phitr}),
Eqs. (\ref{awtr})--(\ref{amutr}) lead to the following 
gauge-invariant quantities
\begin{eqnarray} 
&&
\tilde{{\cal A}}_{w} = \tilde{a}_{w} + (n - P) \biggl[ \frac{M}{L}
\tilde{{\cal P}} - 
\frac{M^2}{L^2} \dot{\tilde{E}}\biggr]',
\label{AW}\\
&& \tilde{{\cal A}}_{\theta} = \tilde{a}_{\theta} 
+ (n - P) \biggl[ \frac{M}{L}\tilde{{\cal P}} - 
\frac{M^2}{L^2} \dot{\tilde{E}}\biggr]^{\cdot} + P' \biggl( 
\frac{M^2}{L^2} \tilde{E}' - \frac{M}{L} \tilde{C}  \biggr),
\\
&& \tilde{\cal A} = \tilde{a} + ( n - P) \biggl[ \frac{M}{L}\tilde{{\cal P}} - 
\frac{M^2}{L^2} \dot{\tilde{E}}\biggr].
\label{Adiv}
\end{eqnarray}
The same type of construction can be carried on in the case of 
the fluctuations of the Higgs field. In this case the 
explicit variation for infinitesimal coordinate transformations 
turns out to be 
\begin{eqnarray}
&& \tilde{\chi} = \chi - \epsilon_{w} \varphi' - \epsilon_{\theta} 
\dot{\varphi},
\nonumber\\
&&\tilde{\chi}^* = \chi^* - \epsilon_{w} {\varphi^*}' - \epsilon_{\theta} 
\dot{\varphi}^*.
\end{eqnarray}
Recalling, again, Eqs. (\ref{El})--(\ref{phitr}), 
the gauge-invariant fluctuation 
corresponding to the Higgs field becomes
\begin{equation}
\tilde{X} = \tilde{\chi} + \biggl( \frac{M}{L} 
\tilde{C} - \frac{M^2}{L^2} \tilde{E}' \biggr) \varphi' 
+ \biggl[ \frac{M}{L}\tilde{{\cal P}} - 
\frac{M^2}{L^2} \dot{\tilde{E}}\biggr]\dot{\varphi},
\label{giH}
\end{equation}
and analogously for the complex conjugate field.
Using Eqs. (\ref{ff2})--(\ref{ff3}), the gauge invariant combination 
corresponding to $g$ will then be, from Eq. (\ref{giH}) 
\begin{equation}
\tilde{\Delta} = \tilde{g} + \biggl( \frac{M}{L} 
\tilde{C} - \frac{M^2}{L^2} \tilde{E}' \biggr) f' 
+ i n \biggl[ \frac{M}{L}\tilde{{\cal P}} - 
\frac{M^2}{L^2} \dot{\tilde{E}}\biggr] f.
\label{giD}
\end{equation}

\subsection{Gauge choices}

Before analyzing the explicit form of the evolution equations for the 
fluctuations
it is appropriate to mention that specific gauge choices can be made 
in full analogy with what happens in the five-dimensional case 
\cite{mg1,mg2}. In spite of the fact that our analysis will be 
gauge-invariant, the physical interpretation of a given result 
is more transparent in a specific gauge.
Of particular relevance is the {\em longitudinal gauge}
where the physical interpretation of the gauge-invariant fluctuations 
becomes particularly simple
\begin{equation}
\tilde{E} =0, ~~~\tilde{P} =0,~~~\tilde{C}= 0, ~~~\tilde{f}_{\mu}=0.
\label{lon1}
\end{equation}
By solving Eqs. (\ref{lon1}) with respect to $\epsilon$, $\epsilon_{w}$, 
$\epsilon_{\theta}$ and $\zeta_{\mu}$, an by recalling Eqs. (\ref{fl})--(\ref{zeta2}) and 
(\ref{El})--(\ref{phitr}) we get 
\begin{eqnarray}
&&\epsilon = E ,
\nonumber\\
&& \epsilon_{w} = \biggl( \frac{M^2}{L^2} E' - \frac{M}{L} C\biggr),
\nonumber\\
&& \epsilon_{\theta} = \frac{M^2}{L^2} \dot{E} -\frac{M}{L} P,
\nonumber\\
&&\zeta_{\mu} = f_{\mu}.
\label{lon2}
\end{eqnarray}
This gauge fixing was called {\em longitudinal} \cite{mg1,mg2} since 
the off-diagonal 
(scalar) fluctuations vanish and the perturbed form of the metric given in Eq. 
(\ref{lorf}) contains, in its off-diagonal entries, only 
pure vector fluctuations. 
Interestingly enough, using Eqs. (\ref{lon1})--(\ref{lon2}) 
we have that, in this
gauge, the gauge-invariant fluctuations defined above greatly simplify.
For the vectors we have that
\begin{equation}
V_{\mu} = D_{\mu},~~~~Z_{\mu} = Q_{\mu},
\end{equation}
whereas for the scalars 
\begin{equation}
\Psi = \psi, ~~~~\Phi= \phi,~~~~\Xi= \xi, ~~~~\Pi= \pi,
\end{equation}
and similarly for the sources. This shows, as stated before, that in the 
longitudinal gauge, the off-diagonal elements of the metric are pure vectors.

\renewcommand{\theequation}{4.\arabic{equation}}
\setcounter{equation}{0}
\section{Evolution equations for the fluctuations}
Denoting with $\delta$ the first order fluctuation of the corresponding 
tensor, 
the perturbed Einstein equations can be written as 
\begin{equation}
\delta R_{A B} = \kappa \delta \tau_{A B} ,
\label{einper}
\end{equation}
where  
\begin{equation}
\delta R_{AB} = \partial_{C} \delta \Gamma_{A B}^{C} - \partial_{B} \delta 
\Gamma_{A C}^{C} + \overline{\Gamma}_{A B}^{C} \delta \Gamma_{ C D}^{D} 
+  \delta\Gamma_{A B}^{C} \overline{\Gamma}_{ C D}^{D}- 
\delta\Gamma_{B C}^{D} \overline{\Gamma}_{A D}^{C} - 
\overline{\Gamma}_{B C}^{D} \delta\Gamma_{A D}^{C}. 
\label{ricci}
\end{equation}
In Eq. (\ref{ricci}), $\overline{\Gamma}_{ A B}^{C}$ are the background 
values of the Christoffel connections [computed from Eq. (\ref{newm})]
and $\delta \Gamma_{A B}^{C}$ their first order fluctuations. In Appendix A 
all the explicit form of the Ricci fluctuations in the 
perturbed metric (\ref{lorf})  
are reported [see Eqs. (\ref{munu})--(\ref{thetwp})]
 together with the explicit expressions of
the perturbed Christoffel connection [see Eqs. (\ref{deltach})]
and  together with the perturbed components of the 
 energy-momentum tensor [see Eqs. (\ref{tmn})--(\ref{tthw})] whose
general expression is  
\begin{eqnarray}
\delta \tau_{A B} 
 &=&   \biggl[ - \frac{\Lambda }{2}  + \frac{1}{8} F_{M N}F^{ M N} 
- \frac{\lambda }{8} ( \varphi^* \varphi - v^2)^2\biggr] \delta G_{A B} 
\nonumber\\
&+&  \frac{i e}{2} \delta A_{A} ( \varphi^* \partial_{B}\varphi  - \varphi 
\partial_{B}\varphi^*) 
+ \frac{i e}{2} \delta A_{B} ( \varphi^* \partial_{A}\varphi  - \varphi 
\partial_{A}\varphi^*) 
\nonumber\\
&+&  \frac{i e}{2}  A_{A} ( \chi^* \partial_{B}\varphi+ 
\varphi^* \partial_{B}\chi  - \chi \partial_{B}\varphi^*    - \varphi 
\partial_{B}\chi^*)
\nonumber\\   
&+&  \frac{i e}{2}  A_{B} ( \chi^* \partial_{A}\varphi+ 
\varphi^* \partial_{A}\chi  - \chi \partial_{A}\varphi^*    - \varphi 
\partial_{A}\chi^*)
\nonumber\\
&+& e^2 \delta A_{A } A _{B} \varphi^* \varphi + 
e^2 A_{A } \delta 
A _{B} \varphi^* \varphi + e^2 A_{A } A _{B} 
( \chi^* \varphi + \varphi^* \chi)
\nonumber\\
&-& \delta F_{A C} F_{B D} G^{D C} - F_{A C} \delta F_{ B D} G^{D C} 
- F_{A C} F_{B D} \delta G^{D C}  
\nonumber\\
&+&\frac{G_{A B}}{8}\biggl[
 \delta F_{M N} F_{C D} G^{ C M} G^{D N} +
  F_{M N} \delta F_{C D} G^{ C M} G^{D N} +  
 F_{M N} F_{C D} \delta G^{ C M} G^{D N} 
\nonumber\\
&+&
 F_{M N} F_{C D}  G^{ C M} \delta G^{D N}\biggr]
- \frac{\lambda}{4} ( \varphi^* \varphi - v^2) ( \chi \varphi^* + 
\varphi \chi^*) G_{A B}
\nonumber\\
&+& \frac{1}{2} \left[\partial_A \chi^*\partial_{B}\varphi +
\partial_A \varphi^*\partial_{B}\chi 
+\partial_B\chi^*\partial_A\varphi +
\partial_B\varphi^*\partial_A\chi  \right],
\label{tauper}
\end{eqnarray}
where $\delta F_{C D} = \partial_{C} \delta A_{D} - \partial_{D} \delta A_{C}$.

Eq. (\ref{einper}) is supplemented by 
the perturbed version of Eqs. (\ref{ph})--(\ref{A}). The 
first order fluctuation of Eq. (\ref{ph}) gives
\begin{eqnarray}
&&\delta G^{A B}\biggl( \partial_{A} \partial_{B} \varphi 
- \overline{\Gamma}_{A B }^{C} \partial_{C} \varphi\biggr)
\nonumber\\
&& G^{A B}\biggl(  \partial_{A} \partial_{B} \chi 
- \overline{\Gamma}_{A B }^{C} \partial_{C} \chi -
 \delta\Gamma_{A B }^{C} \partial_{C} \varphi \biggr) -e^2 \delta G^{A B}
A_{A} A_{B} \varphi
\nonumber\\
&& - e^2 G^{A B} \delta A_{A} A_{B} \varphi - e^2 G^{A B} A_{A} 
\delta A_{B} \varphi - 
e^2 G^{A B} A_{A} A_{B} \chi 
\nonumber\\
&& - ie \delta \delta G^{ A B} A_{A} \partial_{B} \varphi - i e 
G^{A B} A_{A} \partial_{B} \chi - i e G^{AB} \delta A_{A} \partial_{B} \varphi
\nonumber\\
&&- i e \delta G^{A B}\biggl[ \partial_{A} ( A_{B} \varphi) -
 \overline{\Gamma}_{A B}^{C} A_{C} \varphi \biggr]
\nonumber\\
&& - i e G^{A B} \biggl[\partial_{A} ( \delta A_{B} \varphi) + \partial_{A}
( A_{B} \chi) - \delta \Gamma_{A B}^{C} A_{C} \varphi - 
\overline{\Gamma}_{A B}^{C} \delta A_{C} \varphi - \overline{\Gamma}_{A B}^{C} 
A_{C} \chi\biggr]
\nonumber\\
&&+ \lambda( \varphi^* \varphi - v^2) \chi + \lambda( \varphi^* \chi + 
\varphi \chi^*)\varphi =0.
\label{phper}
\end{eqnarray}
Finally, the first order fluctuation of Eq. (\ref{A}) leads to 
\begin{eqnarray}
&& \delta G^{ A C} \biggl[ \partial_{C} F_{A B} - \overline{\Gamma}_{ A C}^{D}
 F_{D B} - \Gamma^{D}_{B C} F_{A D}\biggr] + e^2 \delta A_{B} \varphi^* \varphi + e^2 A_{B} 
\biggl[ \chi^* \varphi + \varphi^* \chi \biggr]
\nonumber\\
&& G^{ AC}[ \partial_{C} \delta F_{A B} - \delta \Gamma_{ A C}^{D} F_{D B} - 
\Gamma_{A C}^{D} \delta F_{D B} - \delta \Gamma_{ B C}^{D} F_{ A D} 
- \Gamma_{B C}^{D} \delta F_{AD} \biggr]
\nonumber\\
&& - \frac{ i e}{2} [ \chi \partial_{B} \varphi^* 
+ \varphi \partial_{B} \chi^* - \chi^* \partial_{B} \varphi 
- \varphi^* \partial_{B} \chi] =0.
\label{Aper}
\end{eqnarray}
Three separate gauge-invariant problems then emerge naturally 
within the construction developed up to now. 
Depending upon the transformation properties of the given fluctuation, 
 {\em the tensor problem} involves the 
evolution equations for the {\em spin two fields} 
coming from the geometry, namely 
$h_{\mu\nu}$. {\em The vector problem} involves all the 
pure vectors coming both from the 
geometry and from the fluctuations of the gauge fields, i.e. all the 
{\em spin one fields}.
The relevant gauge-invariant degrees of freedom are, in this case, 
defined according to Eqs. (\ref{V})--(\ref{Z}) and (\ref{amutr})
\begin{equation}
V_{\mu},~~~ Z_{\mu},~~~{\cal A}_{\mu}. 
\end{equation}
Finally,  {\em the scalar problem} involves all the gauge 
invariant degrees of freedom {\em carrying spin zero}, namely, according to Eqs. (\ref{PSI})--(\ref{scgi}), 
(\ref{AW})--(\ref{Adiv}) and (\ref{giD})
\begin{equation}
\Psi,~~~\Phi,~~~\Pi,~~~\Xi,~~~{\cal A}_{w},~~~{\cal A}_{\theta},~~~
{\cal A},~~~\Delta,~~~ \Delta^*.
\end{equation}
Since no pure tensor source is present 
in the Abelian vortex, the tensor fluctuations will be 
decoupled from the very beginning. The gauge-invariant vector 
fluctuations of the geometry will be coupled with the 
vector fluctuations of the vortex. Finally, 
the gauge-invariant scalar fluctuations will mix both with 
the fluctuations of the Higgs and with the scalar fluctuations coming from
the gauge sector.

\renewcommand{\theequation}{5.\arabic{equation}}
\setcounter{equation}{0}
\section{Spin two fluctuations: the tensor problem} 
The explicit expression for the evolution equation of $h_{\mu\nu}$ is
\begin{equation}
\ddot{h}_{\mu\nu} + h_{\mu\nu}'' + 4 {\cal H} h_{\mu\nu} - \frac{L^2}{M^2}
\partial_{\alpha}\partial^{\alpha} h_{\mu\nu} =0,
\label{tens2} 
\end{equation}
is obtained from  the tensor part 
component of Eq. (\ref{einper})
\begin{equation}
\delta R_{\mu \nu} = \kappa \delta \tau_{\mu\nu}
\end{equation}
with the use of the explicit expressions of the Ricci tensors of 
Appendix A and of the background equations of (\ref{e1})--(\ref{e3}) 
(allowing to eliminate the dependence upon the rescaled cosmological constant).

For the zero mode, the solution of Eq. (\ref{tens2}) is, 
for each polarization, 
\begin{equation}
h = K, 
\end{equation}
where $K$ is an arbitrary constant. 
By looking at the normalization of the kinetic term of $h_{\mu\nu}$ in the 
action, the canonical fluctuation is exactly
\begin{equation}
v_{\mu\nu} = M L h_{\mu\nu}.
\end{equation}
Hence, the normalization condition 
for the canonical zero mode is 
\begin{equation}
K^2 \int_{-\infty}^{+\infty}M^2(w)L^2(w) dw. 
\label{normt}
\end{equation} 
But the four-dimensional Planck mass is finite and 
the integral  
\begin{equation}
M_{P}^2 \sim \frac{M_{6}^4}{m_{H}^2} \int_{0}^{\infty} dx M^2(x) {\cal L}(x) \equiv 
\frac{M_{6}^4}{2} \int_{-\infty}^{+\infty}  M^2(w) L^2(w)dw
\end{equation}
is always convergent both for $w \to -\infty$ (going as 
$e^{2 w}$) and for $w\to +\infty$ going as $ (c w)^{-4}$. Since the integral
defining the normalization of the tensor zero mode is the same 
integral appearing in the expression of the four-dimensional 
Planck mass, we can conclude that the tensor zero mode is always 
localized. Notice that in deriving this conclusion no specific 
vortex solution has been used, but only the background equations 
together with the asymptotics which are common to the whole 
class of vortex backgrounds discussed in Section II.

\renewcommand{\theequation}{6.\arabic{equation}}
\setcounter{equation}{0}
\section{Spin one fluctuations: the vector problem} 
From Eq. (\ref{Aper}), taking into account Eq. (\ref{deltach}),
 the pure vector component  of the perturbed gauge 
field equation  leads to the following gauge-invariant expression
\begin{equation}
\frac{L^2}{M^2} \partial_{\alpha} \partial^{\alpha} 
{\cal A}_{\mu} - \ddot{{\cal A}}_{\mu} - {\cal A}_{\mu}'' - 
2 {\cal H} {\cal A}_{\mu}' + 
P'\frac{M}{L}[ Z_{\mu}' -  ({\cal H} - {\cal F}) Z_{\mu} - \dot{V}_{\mu} ]
+ \alpha {\cal A}_{\mu} {\cal L}^2 f^2 =0,
\label{v0}
\end{equation}
where Eqs. (\ref{V})--(\ref{Z}) have been used.

From Eqs. (\ref{einper}), the equations mixing the
vectors coming from the metric and the vector coming 
from the source are
\begin{eqnarray}
&&\delta R_{\mu\nu} = \kappa \delta \tau_{\mu\nu},
\label{pv1}\\
&&\delta R_{\mu w} = \kappa \delta \tau_{\mu w},
\label{pv2}\\
&&\delta R_{\mu \theta} = \kappa \delta \tau_{\mu \theta}.
\label{pv3}
\end{eqnarray}
Using Eqs. (\ref{V}) and (\ref{Z}) into Eqs. (\ref{pv1})--(\ref{pv3})
and bearing in mind the vector part of Eqs. (\ref{munu})--(\ref{muth}) 
and (\ref{tmn})--(\ref{tmth}) 
the following equations are obtained
\begin{eqnarray}
&& V_{\mu}' + ( 3 {\cal H} + {\cal F}) 
V_{\mu} + \dot{Z}_{\mu} =0,
\label{v1}\\
&& 
\frac{M}{2 L} \biggl[\ddot{V}_{\mu} - \frac{L^2}{M^2} 
\partial_{\alpha} \partial^{\alpha} V_{\mu} - \dot{Z}_{\mu}' +
({\cal H} - {\cal F}) \dot{Z}_{\mu}\biggr] -
 \frac{ \nu P'}{\alpha {\cal L}^2} \dot{ {\cal A}}_{\mu} =0,
\label{v2}\\
&&  Z_{\mu}'' + 4{\cal H} Z_{\mu}'+ ({\cal F}' - {\cal H}' + 6 {\cal H} {\cal F} - 5 {\cal H}^2 - {\cal F}^2) 
Z_{\mu} -\frac{L^2}{ M^2} \partial_{\alpha} \partial^{\alpha} Z_{\mu}
\nonumber\\
&-& [ \dot{V}_{\mu}' + \dot{V}_{\mu} ( 5 {\cal H} -{\cal F})]
+2 \frac{ L}{M}\biggl[
\frac{\nu P'}{\alpha {\cal L}^2} {\cal A}_{\mu}' + \nu P f^2
{\cal A}_{\mu}\biggr]
 =0.
\label{v3}
\end{eqnarray}

Eqs. (\ref{v0}) together with   
(\ref{v1})--(\ref{v3}) form a system determining the 
coupled evolution of the vector fluctuations coming from the 
gauge and metric sector.In order to find the zero modes of the system, 
define first, the background function 
\begin{equation}
\varepsilon= \frac{ L}{M},
\end{equation}
whose derivatives satisfy 
\begin{equation}
\frac{\varepsilon '}{\varepsilon}  = ({\cal F} - {\cal H} ) = 
- \frac{\nu}{\alpha} \frac{P~P'}{{\cal L}^2}.
\end{equation}
as dictated by Eqs. (\ref{hminf}) and as a consequence of the 
relations among the string tensions.
Define then,  
the following  combination of the gauge-invariant graviphoton fields
\begin{equation}
\mu_{\alpha} =\varepsilon  \dot{V}_{\alpha} - (\varepsilon Z_{\alpha})'.
\label{combv}
\end{equation} 
Inserting now Eq. (\ref{combv})  
into Eq. (\ref{v0}) and  
 Eqs. (\ref{v2})--(\ref{v3}), the system becomes 
\begin{eqnarray}
&&\ddot{{\cal A}}_{\alpha} + {\cal A}_{\alpha}'' + 2 {\cal H} {\cal A}'_{\alpha} 
- \varepsilon^{2} \Box {\cal A}_{\alpha} + 
\frac{P'}{\varepsilon^2}  \mu_{\alpha} - \alpha {\cal L}^2 f^2 
{\cal A}_{\alpha} =0,
\label{v0a}\\ 
&& \dot{\mu}_{\alpha} -
 \varepsilon^3 \Box V_{\alpha} 
= 2 \varepsilon^2 \frac{\nu}{\alpha} \frac{P'}{{\cal L}^2} 
\dot{{\cal A}}_{\alpha},
\label{v2a}\\
&& \mu_{\alpha}' + ( 4 {\cal H} - 2 \frac{\varepsilon '}{\varepsilon}) 
\mu_{\alpha} + 
\varepsilon^3 \Box  Z_{\alpha} =
2 \varepsilon^2 \biggl[ \nu P f^2 {\cal A}_{\alpha} 
+ \frac{\nu}{\alpha} \frac{P'}{{\cal L}^2} {\cal A}_{\alpha}' \biggr],
\label{v3a}
\end{eqnarray}
subjected to the constraint 
\begin{equation}
V'_{\alpha} + \biggl( 4 {\cal H} + \frac{\varepsilon'}{\varepsilon}\biggr) 
V_{\alpha} + \dot{Z}_{\alpha} =0.
\label{con2}
\end{equation}

Thanks to four-dimensional Poincar\'e invariance the four-dimensional 
D'Alembertian can be replaced by $-m^2$ where $m$ denotes the mass eigenvalue
of the corresponding fluctuations.
Furthermore,  a generic fluctuation can be 
expanded in a Fourier series in $\theta$. For instance
\begin{equation}
V_{\mu}(x^{\mu},w,\theta) = \sum_{\ell=-\infty}^{\infty} V_{\mu}^{(\ell)}
(x^{\mu},w) 
e^{i \ell \theta}.
\end{equation}
Consider now the situation where $V_{\mu}$ is massless, i.e.
$\Box V_{\mu} =0$. From Eq. (\ref{v2a})
\begin{equation}
\mu_{\alpha} = 2 \varepsilon^2 \frac{\nu}{\alpha} 
\frac{ P'}{{\cal L}^2} {\cal A}_{\alpha}. 
\label{mua}
\end{equation}
Inserting Eq. (\ref{mua}) into Eq. (\ref{v0a}) the following 
decoupled equation is easily obtained by using, simultaneously,  
Eq. (\ref{Pg}) together with Eq. (\ref{hminf})
\begin{equation}
\ddot{{\cal A}} + {\cal A}_{\alpha}'' + 2 {\cal H} {\cal A}'_{\alpha} 
- \varepsilon^{2} \Box {\cal A}_{\alpha} - 
\biggl[ \frac{P''}{P} + 2 {\cal H}\frac{P'}{P} \biggr]
{\cal A}_{\alpha} =0,
\end{equation}
which can also be written as 
\begin{equation}
{\cal N}_{\alpha}'' - \biggl[\frac{(M P)''}{MP}\biggr] {\cal N}_{\alpha} 
+\biggl[  \ddot{\cal N}_{\alpha} - \biggl(\frac{L}{M}\biggr)^2 
\Box {\cal N}_{\alpha}  \biggr] =0,
\label{BA}
\end{equation}
where ${\cal N }_{\alpha} = M {\cal A}_{\alpha} $. 
Eq. (\ref{BA}) holds 
for any mass or angular momentum eigenstates of the vector 
${\cal A}_{\alpha}$. 
Inserting now Eq. (\ref{mua}) into Eq. (\ref{v3a})  and using 
the background equations we get 
\begin{equation}
\Box Z_{\alpha } =0,
\end{equation}
implying that also $Z_{\alpha}$ should be massless. 

From Eq. (\ref{BA})  the solution for the zero mode can be obtained
\begin{equation}
{\cal A}_{\alpha} = K_{1,\alpha}  P(w) + K_{2,\alpha} P(w) 
\int_{-\infty}^{w} \frac{ d w'}{M(w')^2 P(w')^2}.
\label{0ma}
\end{equation}
Since $M(w')P(w') \to 1$ for $w \to -\infty$, the integral appearing 
in the second solution diverges in the lower limit of integration. 
The gauge zero mode is then given by 
\begin{equation}
{\cal A}_{\alpha} = K_{1,\alpha} P(w). 
\label{01ma}
\end{equation}
In order to check for the normalizability of the zero mode 
the gauge action has to be perturbed to second order. By doing so we find that 
the corresponding kinetic term appears in the action for the fluctuations as 
\begin{equation}
\int d^4x~dw~d\theta [L^2 \partial_{\mu} 
{\cal A}_{\alpha} \partial_{\nu} {\cal A}_{\beta} 
\eta^{\mu \nu} \eta^{\alpha\beta}].
\end{equation} 
Hence, the canonical zero mode is $K_{1,\alpha} L(w) P(w)$
and the normalization condition reads 
\begin{equation}
K_{1,\alpha}^2 \int_{-\infty}^{+\infty} L^2(w) P(w)^2 =1.
\label{norm1}
\end{equation}
From Eqs. (\ref{w-inf})--(\ref{Pw}), 
the asymptotic behavior of the integrand of Eq. (\ref{norm1})  
in the two limits of integration is , respectively, 
\begin{eqnarray}
&& L(w)^2 P(w)^2 \simeq e^{2 w},~~~~~w\to -\infty,
\nonumber\\
&& L(w)^2 Pw)^2 \simeq ( c w)^{- 2 - 2 \gamma_{P}}, ~~~~~w\to + \infty.
\end{eqnarray}
Since $\gamma_{P}\gg 1$, the integral converges in both limits and the gauge 
zero mode is normalized \footnote{According to \cite{mgv1}, localized gauge zero modes 
are present in five-dimensions, provided the four-dimensional 
Planck mass is {\em not} finite. In the present case, the localization of the gauge zero mode 
occurs in a six-dimensional geometry (leading to a {\em finite} four-dimensional Planck mass) and
 {\em in the presence } of a gauge field background (which is absent 
in the case of \cite{mgv1}).} . 
Furthermore, the appropriate boundary conditions 
for the zero mode are satisfied. In fact, from Eq. (\ref{01ma}) 
\begin{equation}
{\cal A}_{\alpha}'(-\infty) = {\cal A}_{\alpha}'(+\infty) =0.
\end{equation}

Consider now the zero modes of the gauge-invariant vector fluctuations of the 
geometry, namely $V_{\mu}$ and $Z_{\alpha}$. For the lowest 
angular momentum eigenstates $\dot{V}_{\mu} =0$ and
$\dot{Z}_{\alpha} =0$. Hence, from Eq. (\ref{con2}) 
the zero mode of $V_{\mu}$ is simply obtained and it is 
\begin{equation}
V_{\alpha}(w) = \frac{{\cal C}_{1,\alpha}}{M(w)^3 L(w)},
\label{0mv}
\end{equation}
where ${\cal C}_{1,\alpha}$ is the integration constant.
Using Eq. (\ref{01ma})  
into Eq. (\ref{mua}) the zero mode of $Z_{\alpha}$
is obtained recalling Eq. (\ref{combv}):
\begin{equation}
Z_{\alpha}(w) = K_{1,\alpha} \frac{L}{M} + 
{\cal C}_{2,\alpha} \frac{M}{L},
\label{0mz}
\end{equation}
where ${\cal C}_{2,\alpha}$ is the further integration constant
while $K_{1,\alpha}$ is the same constant appearing in Eq. (\ref{01ma}) and which 
is determined from Eq. (\ref{norm1}).

In order to assess the localization of the vector fluctuations 
of the metric,  the appropriate normalization of the kinetic term of the fluctuations 
has to be deduced by perturbing the action to second  order. 
It is better to perturb to second order 
the Einstein-Hilbert action  directly in the form 
\begin{equation}
G^{A B}\biggl( \Gamma_{A C}^{D} \Gamma_{B D}^{C}
- \Gamma_{A B}^{C} \Gamma_{C D}^{D}\biggr) 
\end{equation}
where the total derivatives are absent.
The kinetic terms appear in the action as 
\begin{eqnarray}
&& \int d^4 x ~ dw~ d\theta M^2 L^2 [ \partial_{\alpha} Z_{\mu} 
\partial_{\beta} Z_{\nu} \eta^{\alpha\beta} \eta^{\mu\nu}],
\nonumber\\
&& \int d^4 x ~ dw ~d\theta M^2 L^2 [ \partial_{\alpha} V_{\mu} 
\partial_{\beta} V_{\nu} \eta^{\alpha\beta} \eta^{\mu\nu}]
\end{eqnarray}
From Eqs. (\ref{0mv}) and (\ref{0mz}) 
the normalization condition for $V_{\mu}$ and $Z_{\mu}$ 
lead, respectively, to the following two integrals
\begin{equation}
{\cal C}_{1}^2 \int_{-\infty}^{\infty} \frac{ d w}{M(w)^4},
\label{norm2}
\end{equation}
and 
\begin{equation}
\int_{-\infty}^{\infty} \biggl[ K_{1,\alpha} L(w)^2 + {\cal C}_{2,\alpha} 
M(w)^2\biggr]^2  ~~dw.
\label{norm3}
\end{equation}
For $w\to +\infty$, $ M(w)^{-4} \to (c w)^4$, as implied by the ${\rm AdS}_{6}$ 
nature of the geometry, in this limit. The  normalization 
integral of Eq. (\ref{norm2}),
corresponding to the $V_{\mu}$ fluctuation, diverges at 
infinity. Consequently the zero mode of $V_{\mu}$ is not localized. 

In Eq. (\ref{norm3}) $K_{1,\alpha}$ is determined from the 
normalization of the gauge zero mode according to Eq. (\ref{norm1}).
The integrand appearing in Eq. (\ref{norm3}) 
has three terms: two going, respectively,  as $L^4$ and $M^2 L^2$ and one going 
as $M^4$. The term going as $M^4$ makes the integral divergent. 
For $w \to -\infty$,  $M(w)^2 \to 1$ and, hence, the integral
appearing in Eq. (\ref{norm3}) will be linearly divergent for 
$w \to -\infty$.  

The situation can then be summarized by saying that while the gauge zero 
mode is localized, the vector modes of the geometry are never localized. 
One of them, $V_{\mu}$,  because of the behavior at infinity, the
other, $Z_{\mu}$, because of the behavior in the core of the defect.
It is interesting, in this context, to repeat the calculation described in the 
present Section in the coordinate system defined by the line element of Eq. (\ref{metric}).
In this case  the evolution equations of the fluctuations 
are less symmetric but, in spite of this, the same conclusions can be reached 
through a slightly different algebraic procedure. The results of this 
exercise are reported in Appendix B.

Two final remarks are in order.
To get to the correct conclusion is crucial to discuss 
the full system describing the fluctuations of the geometry together with 
the fluctuation of the source. Consider then the case where
$P' \to 0$, $P \to 1$ 
 and the the gauge field fluctuations
are absent, i. e. ${\cal A}_{\alpha} =0$. 
If $P' \to 0$, then, 
$\varepsilon \to 1$ and ${\cal H} \to {\cal F}$. In this case 
the space-time is ${\rm AdS}_{6}$ and the evolution of the fluctuations 
follow from Eqs. (\ref{v1})--(\ref{v3}):
\begin{eqnarray}
&& V_{\mu}' + 4 {\cal H} V_{\mu} + \dot{Z}_{\mu} =0,
\label{v1c}\\
&& \ddot{V}_{\mu} - \Box V_{\mu} - \dot{Z}_{\mu}' =0
\label{v2c}\\
&& Z_{\mu}'' + 4 {\cal H} Z_{\mu}' - \Box Z_{\mu} -
[ \dot{V}_{\mu}' + 4 {\cal H} \dot{V}_{\mu} ] =0.
\label{v3c}
\end{eqnarray}
Using Eq. (\ref{v1c}) into Eq. (\ref{v3c}) the following decoupled equation
can be obtained
\begin{equation}
\ddot{Z}_{\mu} + Z_{\mu}'' - \Box Z_{\mu} + 4 {\cal H} Z_{\mu}' =0,
\end{equation}
which implies that the canonically normalized 
zero mode, i.e. $M^2 Z_{\mu}$, 
 behaves as $ M(w)^2$. 
The canonically normalized zero mode related to $V_{\mu}$, 
i.e. $M^2 V_{\mu}$ behaves, on the contrary, as $M(w)^{-2}$. Therefore, 
we would erroneously conclude that while $V_{\mu}$ is not normalizable,
$Z_{\mu}$ is normalizable. This argument overlooks the behaviour of 
the wavefunctions close to the core of the vortex which implies 
that also $Z_{\mu}$ is not normalizable. 

Eq. (\ref{BA}) can be written as 
\begin{equation}
{\cal N}_{p,\ell}'' + \biggl[ m^2 -\ell^2 - \frac{(M P)''}{M P}\biggr] {\cal N}_{p,\ell}=0,
\label{exeq}
\end{equation}
where the polarization index has been suppressed in order to leave room for the 
eigenvalues indices. Eq. (\ref{exeq}) can be studied using the 
usual techniques of supersymmetric quantum mechanics \cite{susyqm} by associating 
the appropriate superpotential to the Schr\"odinger-like potential 
$(M P)''/(M P)$. Since the effective potential goes to zero for 
$w\to \infty$ the spectrum 
will probably be continuous. Moreover, stability requires that $m> \ell$. Finally using
the explicit solutions of Eq. (\ref{exeq}) in the different physical limits the massive 
eigenstates of the system can be analyzed by inserting 
the obtained solutions back into Eq. (\ref{v0a})--(\ref{v3a}).

\renewcommand{\theequation}{7.\arabic{equation}}
\setcounter{equation}{0}
\section{Spin zero fluctuations : the scalar problem} 

The gauge-invariant metric fluctuations 
of spin zero couple both to the scalar fluctuations 
generated by the Higgs field and to the scalar fluctuations coming from the 
gauge field. 
Eq. (\ref{phper}) leads, respectively, to the following  expressions for the real and 
imaginary parts of the Higgs fluctuations:
\begin{eqnarray}
&& \varepsilon^2 \Box \Delta_1 - \Delta_1''- 
\ddot{\Delta}_1  - 4{\cal H}\Delta_1'
+ \Big[P^2+{\cal L}^2(2f^2-1) \Big]\Delta_1 
\nonumber\\
&+& {\cal L}^2 f^2 \Delta_1  - 4f' \Psi' +2P^2f\Phi + f' \Phi'-
  2(f''+4{\cal H} f') \Xi -f' \Xi'
\nonumber\\
&-&  f' \dot{\Pi}  - 2Pf{\cal A}_\theta  
+ 2 P \dot{\Delta}_2 =0,
\label{real}
\end{eqnarray}
and 
\begin{eqnarray}
&& \varepsilon^2 \Box \Delta_2 - \Delta_2''- 
\ddot{\Delta}_2  - 4{\cal H}\Delta_2'
+ \Big(P^2+{\cal L}^2(2f^2-1) \Big)\Delta_2
\nonumber\\
&-&  {\cal L}^2 f^2 \Delta_2 - 4 P f \dot{\Psi} + P f
( \dot{\Xi} -  \dot{\Phi} ) - ( 4 {\cal H} P f + P' f + 2 f' P)\Pi 
- f P \Pi' 
\nonumber\\
&+& f \biggl[- \varepsilon^2 \partial_{\alpha}\partial^{\alpha}{\cal A} + 
{\cal A}_{w}' + \dot{{\cal A}}_{\theta}\biggr] + ( 2 f' + 4{\cal H} f) 
{\cal A}_{w} =0,
\label{im}
\end{eqnarray}
where, according to the notations of Eq. (\ref{giD}) we wrote 
\begin{equation}
\Delta = \Delta_1 + i \Delta_2.
\end{equation}
From the scalar components of Eq. (\ref{Aper}) the following set of 
equations is obtained
\begin{equation}
 {\cal A}'' + \ddot{{\cal A}} - {\cal A}'_w - 
\dot{{\cal A}}_{\theta} + 2 {\cal H} ( {\cal A}'-{\cal A}_w)
= \alpha {\cal L}^2 f^2 {\cal A} - \alpha {\cal L}^2 f \Delta_2,
\label{div}
\end{equation}
\begin{eqnarray}
&& \varepsilon^2 \Big[ \Box ({\cal A}_w - 
{\cal A}') \Big]
- \ddot{{\cal A}}_w + \dot{{\cal A}}'_\theta - P'\Big(\dot{\Phi} +
\dot{\Xi} + (4{\cal H} -2{\cal F}) \Pi + 4 \dot{\Psi}  \Big)
\nonumber\\ 
&& -  P'' \Pi + \alpha {\cal L}^2 f^2 {\cal A}_w - 
\alpha {\cal L}^2 \Big(f \Delta_2' - f'\Delta_2 \Big) = 0,
\label{aw}
\end{eqnarray}
\begin{eqnarray}
&& \varepsilon^2 \Box ({\cal A}_\theta -
\dot{{\cal A}}
) + 2 \Big[ P''+2 P'(2{\cal H}-{\cal F}) \Big] \Xi
+ P' \Big(\Phi' + \Xi' + 4\Psi' \Big) 
\nonumber \\ 
&+& (4 {\cal H} - 2 {\cal F} )( \dot{{\cal A}}_w - 
{\cal A}'_\theta) + \dot{{\cal A}}'_w - 
{\cal A}''_\theta 
\nonumber \\
&+& \alpha {\cal L}^2 f^2 {\cal A}_\theta  -2 P \alpha {\cal L}^2 f  
\Delta_1 - \alpha {\cal L}^2 f \Delta_2 =0.
\label{ath}
\end{eqnarray}
Eq. (\ref{div}) represents the divergence-full part of Eq. (\ref{Aper}). 
Eqs. (\ref{aw}) and Eq. (\ref{ath}) are the 
$w$ and $\theta$ component of Eq. (\ref{Aper}).

Using the results of Appendix A, the 
$(\mu \neq \nu)$, $(\mu = \nu)$, $(\mu, w)$ and $(\mu ,\theta)$ components 
of Eq. (\ref{einper}) lead, respectively, to
\begin{eqnarray}
&& \Phi +\Xi - 2\Psi  = 0,
\label{mn}\\
&&\Psi''+ \ddot{\Psi} - \varepsilon^2 \Box \Psi + 8 {\cal H} \Psi' +
{\cal H} \Xi' + 2 ({\cal H}' + 4 {\cal H}^2) \Xi 
 - {\cal H} (\Phi' - \dot{\Pi}) =
\nonumber \\
 && \frac{\nu}{2 \alpha {\cal L}^2} 
\Big[P'(\dot{{\cal A} _w} - {\cal A}_\theta') + 
P'^2 (\Phi+\Xi)\Big] -2 \frac{\nu 
{\cal L}^2}{4} f (f^2-1) \Delta_1,
\label{mn2}\\
&&\Phi' + ({\cal F}-{\cal H}) \Phi - \frac{1}{2} \dot{\Pi} - 
(3{\cal H} + {\cal F}) \Xi - 3 \Psi' =
\nu f' \Delta_1 + \frac{\nu P'}{\alpha {\cal L}^2}(\dot{{\cal A}} - 
{\cal A}_\theta), 
\label{mw}\\
&&\dot{\Xi} - \frac{1}{2} \Pi' - ({\cal H} + {\cal F}) \Pi
-3\dot{\Psi} = \nu f P \Delta_2
+ \frac{\nu P'}{\alpha {\cal L}^2} ({\cal A}_w-{\cal A}')
-\nu f^2P {\cal A}, 
\label{mth}
\end{eqnarray}
whereas the $(w,w)$, $(w,\theta)$ and $(\theta,\theta)$  components
of Eq. (\ref{einper}) give 
\begin{eqnarray}
&-& \varepsilon^2 \Box \Xi + \ddot{\Xi} - (4 {\cal H} + {\cal F}) \Xi' + \Big[ \mu {\cal L}^2 + \frac{\nu 
{\cal L}^2}{4} (f^2-1)^2\Big] \Xi + \Phi'' + {\cal F} \Phi' - 
\frac{3}{2} \frac{P'^2 \nu}{\alpha {\cal L}^2} \Phi 
\nonumber \\
&& - \dot{\Pi}' - 
{\cal F} \dot{\pi}  - 4 \Psi'' + 4 
({\cal F}- 2 {\cal H}) \Psi' = 
2\frac{\nu {\cal L}^2}{4} (f^2-1)f \Delta_1 
\nonumber \\
&+& 2 \nu f' \Delta_1' + \frac{3}{2} 
\frac{\nu P'}{\alpha {\cal L}^2} (\dot{{\cal A}_w} - {\cal A}'_\theta),
\label{ww}\\
&&\frac{1}{2} \varepsilon^2 \Box \Pi + \nu P^2 
f^2 \Pi + 4 {\cal H} \dot{\Xi} 
+ 4 \dot{\Psi}' + 4 ({\cal H} - {\cal F}) \dot{\Psi}  = \nu P f^2 
{\cal A}_w
 \nonumber \\
&-&  \nu f' \Big(\dot{\Delta_1}  - P \dot{\Delta}_2 \Big) 
- \nu P f \Delta_2',
\label{wth}\\
&&- \varepsilon^2 \Box \Phi + \Phi'' + (4 
{\cal H} + {\cal F}) \Phi' - \nu \Big( \frac{3}{2} 
\frac{P'^2}{\alpha {\cal L}^2} + 2 P^2 f^2\Big) \Phi - 
\dot{\Pi}'
\nonumber \\
&-& (4{\cal H} + {\cal F}) \dot{\Pi}
+ \ddot{\Xi} - {\cal F} \Xi' -2 \Big(4{\cal H}{\cal F} + {\cal F}' 
+ \frac{3\nu P'^2}{4\alpha {\cal L}^2} \Big) 
\Xi -4(\ddot{\Psi}+ {\cal F}  \Psi')
\nonumber \\
&=& \frac{3\nu 
P'}{2 \alpha {\cal L}^2} (\dot{{\cal A}}_w - {\cal A}'_\theta) +
 \Big[ \frac{\nu {\cal L}^2}{2} (f^2 - 1)f + P^2 \nu f \Big] \Delta_1 - 2P\nu f^2 
{\cal A}_\theta + 2 \nu P f \dot{\Delta}_2.
\label{thth}
\end{eqnarray}

The lowest angular momentum eigenstates are recovered by setting to zero 
the derivatives with respect to $\theta$. The  system of 
scalar fluctuations  separates then into two
coupled sets of equations. In the first set the off-diagonal scalar 
$\Pi$ couples with the imaginary part of the Higgs field $\Delta_2$ 
together with ${\cal A}$ and  ${\cal A}_{w}$. In the second set 
$\Psi$ and $\Xi$ couple with ${\cal A}_{\theta}$ and with 
the real part of the Higgs fluctuation $\Delta_1$. 

Using the  variables,
\begin{eqnarray}
&& q = {\cal A}' - {\cal A}_{w}, 
\nonumber\\
&& {\cal T} = \frac{\Delta_2}{f} -  {\cal A},
\label{defod}
\end{eqnarray}
 Eqs. (\ref{div})--(\ref{aw}) give 
\begin{eqnarray}
&&\varepsilon^2 \Box q + 2 \nu P f^2 \Big( P \Pi + q + {\cal T}'\Big) =0,
\label{od1}\\
&& q' + 2 {\cal H} q + \alpha {\cal L}^2 f^2 {\cal T} =0.
\label{od2}
\end{eqnarray}
Inserting Eqs. (\ref{defod}) into  Eqs. (\ref{mth}) and (\ref{wth}) we obtain
\begin{eqnarray}
&& \Pi' + 2 ({\cal H} + {\cal F}) \Pi + 2 \nu f^2 P {\cal T} - 2 \frac{\nu}{\alpha} \frac{P'}{{\cal L}^2} q =0,
\label{od3}\\
&&\varepsilon^2 \Box \Pi  +  2 \nu P f^2 \Big( P \Pi + q + {\cal T}'\Big) =0.
\label{od4}
\end{eqnarray}
Finally, inserting Eq. (\ref{defod}) into Eq. (\ref{im}):
\begin{eqnarray}
&& {\cal T}'' + 4 {\cal H} {\cal T}' - [ P^2 + {\cal L}^2 ( f^2 - 1) {\cal T}\bigr] {\cal T} - 
\varepsilon \Box {\cal T} 
\nonumber\\
&& - f q' - 2 ( f' + 2 {\cal H} f) - ( 4 {\cal H} P + P' f + 2 f' P) \Pi =0.
\label{od5}
\end{eqnarray}
Eq. (\ref{od1}) leads to
\begin{equation}
q = - {\cal T}' - P \Pi - \frac{\varepsilon}{ 2 \nu P f^2} \Box q,
\label{od1a}
\end{equation}
and inserting Eq. (\ref{od1a}) into Eq. (\ref{od5}), the following relation holds 
\begin{equation}
\varepsilon^2 \Box\Big( {\cal T} + \frac{ q'}{\alpha {\cal L}^2 f^2} \Big) =0.
\end{equation}
Summing up Eqs. (\ref{od3}) and (\ref{od2})
\begin{equation}
{\cal R}' + 2 ({\cal H} + {\cal F}) {\cal R} =0,
\label{zmod}
\end{equation}
whereas subtracting Eqs. (\ref{od4}) and (\ref{od1}) 
\begin{equation}
\varepsilon^2 \Box {\cal R} =0,
\label{boxod}
\end{equation}
where 
\begin{equation}
{\cal R} = \Pi - \frac{2 \nu P}{ \alpha {\cal L}^2 } q. 
\label{zeta}
\end{equation}
While Eq. (\ref{boxod}) implies that  ${\cal R}$ is a massless 
combination, the solution of 
Eq. (\ref{zmod}) gives the evolution of the zero mode
\begin{equation}
{\cal R}(w) =\Pi - \frac{2 \nu P}{ \alpha {\cal L}^2 }  
\simeq \frac{k_0}{M^2 L^2} 
\label{zet2}
\end{equation}
where $k_0$ is the integration constant. The zero mode 
of $\Pi$ can be obtained by using Eq. (\ref{od1a}) into eq. (\ref{od3}) in the 
case where the mass of $q$ vanishes. 
The obtained relation 
can be written as 
\begin{equation}
\frac{\partial}{\partial w} \Big[ M^4 \Pi + 2 \frac{\nu}{\alpha} P' 
{\cal T} \frac{M^4}{{\cal L}^2}\Big]=0, 
\end{equation} 
whose integral gives 
\begin{equation}
\Pi = - \frac{k_1}{M^4} - 2 \frac{\nu}{\alpha {\cal L}^2} 
P' {\cal T}, 
\label{zeta3}
\end{equation}
where $k_1$ is an integration constant. From Eqs. (\ref{zeta2}) and (\ref{zeta3}), 
$\Pi$ can be eliminated and a relation between $q$ and ${\cal T}$ 
is obtained. Inserting the obtained relation back into Eq. (\ref{od2}) 
the decoupled equation for ${\cal T}$ is found and it can be expressed 
as 
\begin{equation}
\biggl( \frac{L^2}{M^2 P} {\cal T}\biggr)' = {\cal S}(w)
\end{equation}
where 
\begin{equation}
{\cal S}(w) = \frac{\alpha}{2 \nu} \frac{{\cal L}^2 }{P' M^4} \Biggl\{ \frac{k_0}{L^2} \frac{P'}{P} 
+ \frac{k_1}{M^2} \biggl[ 2 ({\cal H} - {\cal F}) + \frac{P'}{P}\biggr]\Biggr\} \simeq
m_{H}^2 \frac{\alpha}{2 \nu P} \frac{k_0 + k_1}{M^2}.
\label{sour}
\end{equation}
The second equality in Eq. (\ref{sour}) follows in te limit where $ M \simeq L$ (i.e. ${\cal H} = 
{\cal F}$), namely in the regime far from the core when the space-time is ${\rm AdS}_{6}$. 
The final expressions for the various zero modes is then 
\begin{eqnarray}
&& {\cal T}(w) = k_2 \frac{M^2}{L^2}  P + \frac{M^2}{L^2}  P \int^{w} dw' {\cal S}(w') ,
\label{zmt}\\
&& q(w) = - k_2 P' \frac{M^2}{L^2} - \frac{\alpha}{2 \nu} \frac{{\cal L}^2}{M^2} 
\Big[ \frac{k_0}{L^2} + \frac{k_1}{M^2}\Big]  - \frac{M^2}{L^2} P' \int^{w} dw' {\cal S}(w') ,
\label{zmq}\\
&& \Pi(w) = - 2 k_2 ( {\cal H} - {\cal F}) \frac{M^2}{L^2} - \frac{k_1}{M^4} + 
\frac{M^2}{L^2} P' \int^{w} dw' {\cal S}(w').
\label{zmp}
\end{eqnarray}
Notice that in the limit of $w\to \infty$, $\Pi$ diverges as $M^{-4} \sim w^4$. 
Since the canonically normalized fluctuation related to $\Pi$ is, for large $w$, 
$M^2 \Pi$, we can also deduce that the gauge-invariant fluctuation $\Pi$ is not 
localized. 

The evolution of the zero modes of $\Phi$ and $\Xi$ is very complicated. 
The cumbersome expressions which can be obtained should 
anyway evaluated in some limit. The zero modes of $\Phi$ and $\Xi$ 
can be obtained by reminding that far from the core of the vortex
the background solutions are determined by the cosmological constant and 
nothing else. In this limit ${\cal H} = {\cal F}$, ${\cal H}^2 = {\cal H}'$ while 
$f= 1 $ and $P =0$. In order to take 
the limit consistently it should also be borne in mind 
that, according to the relations among 
the string tensions, $P P'/{\cal L}^2 \propto ({\cal H} - {\cal F}) \to 0$ .
 The resulting equations are still coupled and they can
be written, for the zero modes, as 
\begin{eqnarray}
&& \Psi'' + 6 {\cal H} \Psi' + 2 {\cal H} \Xi' + 2 ( {\cal H}' + 4 {\cal H}^2) \Xi =0,
\label{dia1}\\
&& \Xi'' + 2 \Psi'' + 2 {\cal H} \Psi' + 2 ( {\cal H}' + 4 {\cal H}^2) \Xi =0,
\label{dia2}\\
&& 2 \Psi'' - \Xi'' + 6 {\cal H} \Psi' - 6{\cal H} \Xi' - 2 ( {\cal H}' + 4 {\cal H}^2) \Xi =0,
\label{dia3}\\
&&\Psi' + \Xi' + 4 {\cal H} \Xi =0.
\label{dia4}
\end{eqnarray}
In order to get Eqs. (\ref{dia1})--(\ref{dia4}), 
 Eq. (\ref{mn}) has been used in order to eliminate $\Phi$ 
in Eqs. (\ref{mn2})--(\ref{mw}) and in Eqs. (\ref{ww})--(\ref{thth}). 

Using Eq. (\ref{dia4}) into Eq. (\ref{dia2}) the following equation for $\Xi$ 
is obtained: 
\begin{equation}
\Xi'' + 4 {\cal H} \Xi' + 6 {\cal H}' \Xi =0,
\end{equation}
implying that that the zero mode goes as  $\Xi \sim M^{-2}$. The canonically
 normalized field is  $M^2 \Xi$. This implies that $\Xi$ is not localized since at 
infinity the normalized zero mode goes as a constant and the 
integral is linearly divergent in $w$ or, exponentially divergent in $x$. Using 
Eq. (\ref{dia4}) the zero mode of $\Psi$ can be also obtained. It corresponds 
to $\Psi \sim {\rm constant}$. In this case the canonical zero mode 
is normalized, at infinity. Finally, from Eq. (\ref{mn}) , it can be 
deduced that $\Phi$ is not normalizable since $\Xi$ is not localized.
The last thing which should be determined is the behaviour 
of $\Psi$ for $w\to -\infty$. In fact $\Psi$ is normalizable at infinity. 
If it would also be normalizable in the origin then, the zero mode of $\Psi$ would be localized. 
The analysis of the equations in the limit $w \to -\infty$ lead to the conclusion
that $\Psi\sim {\rm constant}$ is also a solution. However, since the canonical fluctuation 
related to $\Psi$ is given by $M^2 \Psi$, then also this scalar mode is not localized
In fact, the normalization integral will go as $w$ for $w\to -\infty$.   
\renewcommand{\theequation}{8.\arabic{equation}}
\setcounter{equation}{0}
\section{Concluding remarks}
In this paper the zero modes of 
six-dimensional vortex solutions have been analyzed
in the framework of the gravitating Abelian-Higgs model with a
(negative) cosmological constant in the bulk.
The vortices lead to regular geometries (i.e. free of 
curvature singularities) with finite four-dimensional Planck mass. 
Far from the core of the vortex the geometry is 
determined by the value of the bulk cosmological constant
leading to a ${\rm AdS}_{6}$ space-time.

The analysis of the fluctuations of the vortex 
has been performed in order to check for the localization
of the corresponding zero modes. A given fluctuation is localized
if it is normalizable with respect to the bulk coordinates 
describing the geometry of the vortex in the transverse space.
In order to obtain a realistic low-energy theory 
from higher dimensions it is important to localize 
fields of various spin whose low energy dynamics  could 
lead to the known interactions of the standard model.

The ambiguity deriving from the 
change of the fluctuations for infinitesimal coordinate transformations
around the (fixed) vortex background  
has been resolved by resorting to a fully gauge-invariant approach.
While the tensor zero modes are 
localized on the vortex neither the graviphotons fields (spin one) coming 
from the geometry nor the scalar
fluctuations of the metric are localized. The gauge 
field fluctuation leads to a localized 
zero mode. 

In the model discussed in the present 
paper fermionic degrees of freedom are absent.
It is therefore difficult to understand how the gauge zero mode 
will interact with them. Using recent results \cite{tro1,tro2,lib}
concerning the localization of fermionic zero modes 
on six-diemensional vortices, 
it would be interesting to understand if the gauge zero mode 
discussed in the present paper
could mediate electromagnetic interactions. 
It is in fact unclear if the gauge zero mode is charged 
under an Abelian (local) symmetry which could be interpreted 
as the ordinary electromagnetism. 

In spite of the absence of fermionic 
degrees of freedom in the set-up, the vector field 
coming from the geometry cannot mediate any type 
of electromagnetic interaction. In fact 
it has been shown in general terms that 
the spin one fields associated with 
the fluctuations of the six-dimensional geometry
are not localized on the vortex if 
the four-dimensional Planck mass is finite and if the 
geometry of the vortex is regular.

\section*{Acknowledgments}
The authors are indebted to M. Shaposhnikov 
for interesting discussions  
during different stages of the present work. M.G. would like to acknowledge
the support of the 
Tomalla foundation and the hospitality of the  CERN theory division.
\newpage
\begin{appendix}
\renewcommand{\theequation}{A.\arabic{equation}}
\setcounter{equation}{0}
\section{Explicit expressions of the fluctuations}
Since we ought to obtain gauge-independent evolution equations for the 
fluctuations, it is important to obtain, as a first step, general expressions
for the fluctuations of the Ricci tensors. Subsequently, the evolution 
equations for the fluctuations can be written without specifying the gauge 
but re-expressing the fluctuations in terms of the gauge-invariant 
potentials discussed before. 
The values of the background Christoffel connections are
\begin{eqnarray}
&& \overline{\Gamma}^{w}_{\mu\nu} = \frac{M^2 }{L^2} {\cal H} \eta_{\mu\nu},
\nonumber\\
&& \overline{\Gamma}^{\beta}_{\alpha w} = {\cal H} \delta_{\alpha}^{\beta} ,
\nonumber\\
&& \overline{\Gamma}^{b}_{a w} = {\cal F} \delta_{a}^{b},
\nonumber\\
&& \overline{\Gamma}^{w}_{ a b} = {\cal F} \eta_{a b},
\label{chr}
\end{eqnarray}
[$ a, b$ run over the two transverse dimensions and the Greek indices 
run over the four space-time dimensions]. 

Using the explicit form of the line element 
(\ref{newm}) together with the generic form 
of the perturbed metric given in eq. (\ref{lorf}), 
the eighteen perturbed Christoffel connections 
\begin{eqnarray}
&&\delta \Gamma^{w}_{\mu\nu} = \frac{M^2}{L^2}( H_{\mu\nu}' 
+ 2 {\cal H} H_{\mu\nu}) + 2 \frac{M^2}{L^2} {\cal H} \xi \eta_{\mu\nu} - 
\frac{ M}{2 L} (\partial_{\mu} {\cal G}_{\nu} + \partial_{\nu} {\cal G}_{\mu}),
\nonumber\\
&&\delta \Gamma^{w}_{\mu w} = \frac{M}{L} {\cal H} {\cal G}_{\mu} - 
\partial_{\mu} \xi,
\nonumber\\
&& \delta\Gamma^{w}_{w w} = - \xi',  
\nonumber\\
&&\delta \Gamma^{w}_{\mu\theta} = \frac{M}{2 L}\biggl[ {\cal B}_{\mu}' 
+ ({\cal F}+{\cal H}) {\cal B}_{\mu}\biggr] - \frac{1}{2} \partial_{\mu} \pi 
- \frac{M}{2 L} \dot{ {\cal G}}_{\mu},
\nonumber\\
&& \delta \Gamma^{w}_{w \theta} = - \dot{\xi} + {\cal F} \pi,
\nonumber\\
&& \delta \Gamma^{w}_{\theta \theta} = \phi' - \dot{\pi} + 2 {\cal F} 
( \phi - \xi), 
\nonumber\\
&& \delta \Gamma^{\theta}_{\mu\nu} = \frac{M^2}{L^2} \biggl[ {\cal H} \pi 
\eta_{\mu\nu} + \dot{H}_{\mu\nu}\biggr] - \frac{M}{2L}( \partial_{\mu} 
{\cal B}_{\nu}
 + \partial_{\nu} {\cal B}_{\mu}),
\nonumber\\
&& \delta \Gamma^{\theta}_{\mu\theta} = - \partial_{\mu} \phi,
\nonumber\\
&& \delta \Gamma^{\theta}_{\theta \theta} = - \dot{\phi} - {\cal F}\pi,
\nonumber\\
&& \delta \Gamma^{\theta}_{\mu w} =  \frac{M}{2 L} \biggl(-{\cal B}_{\mu}' 
+ ( {\cal H} - {\cal F} ) {\cal B}_{\mu}\biggr)
+ \frac{M}{2 L} \dot{{\cal G}}_{\mu} -
\frac{1}{2} \partial_{\mu}\pi,
\nonumber\\
&& \delta \Gamma^{\theta}_{\theta w} = - \phi',
\nonumber\\
&& \delta\Gamma^{\theta}_{w w} = \dot{\xi} - \pi' - {\cal F} \pi,
\nonumber\\
&& \delta \Gamma^{\mu}_{\alpha\beta} = - \frac{M}{L} {\cal H} {\cal G}^{\mu} 
\eta_{\alpha\beta} + ( - \partial^{\mu} H_{\alpha\beta} + \partial_{\beta} 
H^{\mu}_{\alpha} + \partial_{\alpha} H^{\mu}_{\beta} ),
\nonumber\\
&& \delta \Gamma^{\mu}_{\alpha w} = {H_{\alpha}^{\mu}}' + \frac{L}{2 M} ( 
\partial_{\alpha} {\cal G}^{\mu} - \partial^{\mu} {\cal G}_{\alpha}),
\nonumber\\
&& \delta\Gamma^{\mu}_{w w} = \frac{L}{M} ( {{\cal G}^{\mu}}' + {\cal H} 
{\cal G}^{\mu}) - \frac{L^2}{M^2} \partial^{\mu} \xi,
\nonumber\\
&& \delta\Gamma^{\mu}_{\theta\theta} = \frac{L}{M} ({\cal F} {\cal G}^{\mu}+ \dot{{\cal B}}^{\mu}) 
- \frac{L^2}{M^2} \partial^{\mu} \phi ,
\nonumber\\
&& \delta \Gamma^{\mu}_{\alpha \theta} = \frac{L}{2 M} \biggl( 
\partial_{\alpha} {\cal B}^{\mu} - \partial^{\mu} {\cal B}_{\alpha} \biggr)
+ \dot{H}_{\alpha}^{\mu}, 
\nonumber\\
&& \delta \Gamma^{\mu}_{\theta w} = \frac{L}{2 M} \biggl( 
{{\cal B}^{\mu}}'+ ({\cal H}- {\cal F}) {\cal B}_{\mu} 
- \partial^{\mu} \pi \biggr) + \frac{L}{2 M} 
\dot{{\cal G}}^{\mu},
\label{deltach}
\end{eqnarray}
lead, through the repeated use of the Palatini identities 
in Eq. (\ref{ricci}) to the following six perturbed Ricci tensors,
\begin{eqnarray}
\delta R_{\mu\nu} &=& \frac{M^2}{L^2}\biggl[ H_{\mu\nu}'' 
+ 4{\cal H} H_{\mu\nu}' + 2 H_{\mu\nu} ({\cal H}' + 4 {\cal H}^2) 
+ \ddot{H}_{\mu\nu} \biggr] - \partial_{\alpha}\partial^{\alpha} H_{\mu\nu}
\nonumber\\
&+& \partial_{\alpha}\partial_{\mu} H^{\alpha}_{\nu} 
+ \partial_{\alpha}\partial_{\nu} H^{\alpha}_{\mu} 
+ \partial_{\nu} \partial^{\alpha} H_{\alpha\mu} 
- \partial_{\nu}\partial_{\alpha} H^{\alpha}_{\mu} 
\nonumber\\
&+& \partial_{\mu} \partial_{\nu} [ \xi + \phi - H_{\alpha}^{\alpha}]
\nonumber\\
&+& \eta_{\mu\nu} \biggl\{ - \frac{M}{L} {\cal H} 
\partial_{\alpha} {\cal G}^{\alpha} +
\frac{M^2}{L^2} \biggl[ {\cal H} \dot{\pi} + 2 \xi ({\cal H}' + 4 {\cal H}^2) 
+ {\cal H} ( {H_{\alpha}^{\alpha}}' + \xi' - \phi') \biggr]\biggr\}
\nonumber\\
&-& \frac{M}{2 L} \biggl[ ( \partial_{\mu} {\cal G}_{\nu} + 
\partial_{\nu} {\cal G}_{\mu} )'+ ( 3 {\cal H} + {\cal F} )   
( \partial_{\mu} {\cal G}_{\nu} + \partial_{\nu} {\cal G}_{\mu} ) \biggr]
\nonumber\\
&-& \frac{M}{2L} ( \partial_{\mu} \dot{\cal B}_{\nu} + 
\partial_{\nu} \dot{\cal B}_{\mu} ),
\label{munu}\\
\delta R_{\mu w} &=& \partial^{\alpha} H_{\mu\alpha}' +
\partial_{\mu} [\phi ( {\cal F} - {\cal H}) + \phi' - \frac{1}{2} \dot{\pi} - 
{H_{\alpha}^{\alpha}}' + \frac{L}{2 M} (\partial_{\alpha} {\cal G}^{\alpha})
- ( {\cal F} + 3 {\cal H} ) \xi] 
\nonumber\\
&+& \frac{ M}{2 L} \ddot{\cal G}_{\mu} - \frac{L}{2M}\partial_{\alpha}
\partial^{\alpha} {\cal G}_{\mu} + 
\frac{M}{L} ( {\cal H}' + 4 {\cal H}^2) {\cal G}_{\mu} 
+ 
\frac{M}{2L}\biggl[ -\dot{{\cal B}}_{\mu}' 
+ ({\cal H} - {\cal F}) \dot{\cal B}_{\mu}\biggr],
\label{muw}\\
\delta R_{\mu\theta} &=& \frac{M}{2L}
\biggl[ {\cal B}_{\mu}'' + 4 {\cal H}  {\cal B}_{\mu}' + 
({\cal H}' + 
{\cal F}' + 3 {\cal H}^2 - {\cal F}^2 + 6 {\cal H}{\cal F}) 
{\cal B}_{\mu} - \frac{L^2 }{ M^2} 
\partial_{\alpha} \partial^{\alpha} {\cal B}_{\mu} \biggr]
\nonumber\\
&+& \frac{ M}{2 L} [ - {\dot{\cal G}_{\mu}}' + \dot{\cal G}_{\mu} 
( {\cal F} - 5 {\cal H}) ]+\partial^{\alpha} \dot{H}_{\mu\alpha} 
\nonumber\\
&+& \partial_{\mu} \biggl[ - \frac{\pi'}{2} + \dot{\xi} - ({\cal H} 
+ {\cal F}) \pi - \dot{H}_{\alpha}^{\alpha}  
+ \frac{L}{2 M} \partial_{\alpha} {\cal B}^{\alpha}\biggr],
\label{muth}\\
\delta R_{w w} &=& \frac{L}{M} \biggl[ (\partial_{\alpha} {\cal G}^{\alpha})'
+ {\cal H} (\partial_{\alpha} {\cal G}^{\alpha}) \biggr]
- \frac{L^2}{M^2} \partial_{\alpha} \partial^{\alpha}\xi 
\nonumber\\
&+& \ddot{\xi} - {\dot{\pi}}' - {\cal F} \dot{\pi} 
\nonumber\\
&+& \phi'' - {H_{\alpha}^{\alpha}}'' + ({\cal F} - 2 {\cal H}) 
{H_{\alpha}^{\alpha}}' - ( {\cal F} + 4 {\cal H} ) \xi' + {\cal F} \phi',
\label{wwp}\\
\delta R_{\theta\theta} &=& \frac{L}{M} (\partial_{\alpha} 
\dot{\cal B}^{\alpha}) + \frac{L}{M} {\cal F} \partial_{\alpha} {\cal G}^{\alpha}
\nonumber\\
&+& \phi'' + \phi' ( 4 {\cal H} + {\cal F}) + 
2 ( {\cal F}' + 4 {\cal H} {\cal F} ) \phi - \frac{L^2}{M^2} 
\partial_{\alpha}\partial^{\alpha} \phi 
\nonumber\\
&+& \ddot{\xi} - 2 ( {\cal F}' + 4 {\cal H} {\cal F}) \xi - {\cal F} \xi'
- \dot{\pi}' - \dot{\pi} ( {\cal F} + 4 {\cal H}) - \ddot{H}_{\alpha}^{\alpha} 
- {\cal F} {H_{\alpha}^{\alpha}}',
\label{ththp}\\
\delta R_{\theta w} &=& \frac{L}{2 M}\biggl[ (\partial_{\alpha} 
{\cal B}^{\alpha})' + ({\cal F} - {\cal H}) \partial_{\alpha} 
{\cal B}^{\alpha}\biggr]
 - \frac{L^2}{2 M^2} \partial_{\alpha} \partial^{\alpha} 
\pi + \frac{L}{2 M} (\partial_{\alpha} \dot{{\cal G}}^{\alpha})
\nonumber\\
&+& ( {\cal F}' + 4 {\cal H} {\cal F} )\pi - 
{\dot{H}_{\alpha}^{\alpha\prime}}
+ \dot{H}_{\alpha}^{\alpha} ( {\cal F} - {\cal H}) - 4 {\cal H} \dot{\xi}.
\label{thetwp}
\end{eqnarray}
From Eq. (\ref{tauper}) and using, again, Eqs. (\ref{deltach}) the explicit 
form of the perturbed energy-momentum tensor is obtained:
\begin{eqnarray}
\kappa \delta \tau_{\mu\nu} &=& 2 \frac{ M^2}{L^2} \biggl[ 
-  \frac{\mu}{2} {\cal L}^2 - 
 \frac{\nu}{8}(f^2 - 1 )^2 {\cal L}^2 
+\frac{\nu}{4} \frac{{P'}^2}{\alpha {\cal L}^2}\biggr] H_{\mu\nu}
\nonumber\\
&+& \frac{M^2}{L^2}\eta_{\mu\nu} \biggl\{ \frac{\nu}{2} \frac{{P'}^2}{\alpha {\cal L}^2} ( \phi + \xi) 
- \frac{ \nu P'}{2 \alpha {\cal L}^2} ( a_{\theta}' - \dot{a}_{w}) 
- \frac{\nu}{4} f ( f^2 -1) {\cal L}^2 ( g + g^*)\biggr\},
\label{tmn}\\
\kappa \delta\tau_{\mu w} &=& \frac{M}{L} {\cal G}_{\mu} \biggl[ 
-  \frac{\mu}{2} {\cal L}^2 - 
 \frac{\nu}{8}(f^2 - 1 )^2 {\cal L}^2 
+\frac{\nu}{4} \frac{{P'}^2}{\alpha {\cal L}^2}\biggr] + \frac{\nu P'}{\alpha 
{\cal L}^2} \dot{{\cal A}}_{\mu} 
\nonumber\\
&+&  \partial_{\mu}\biggl[ \frac{\nu}{2} ( g + g^*) f' 
- \frac{ \nu P' }{\alpha {\cal L}^2} (a_{\theta} -\dot{a}) \biggr],
\label{tmw}\\
\kappa \delta \tau_{\mu\theta} &=& \frac{M}{L}{\cal B}_{\mu} \biggl[ 
-  \frac{\mu}{2} {\cal L}^2 - 
 \frac{\nu}{8}(f^2 - 1 )^2 {\cal L}^2 
+\frac{\nu}{4} \frac{{P'}^2}{\alpha {\cal L}^2}\biggr]
-\frac{ \nu P'}{\alpha {\cal L}^2} {\cal A}_{\mu}' - \nu P f^2 {\cal A}_{\mu} 
\nonumber\\
&+& \partial_{\mu}\biggl[ \frac{i~\nu}{2} P f ( g^* - g) 
+ \frac{\nu P'}{\alpha {\cal L}^2 } (a_{w} - a') - \nu f^2 P a\biggr],
\label{tmth}\\
\kappa\delta \tau_{w w} &=& 2 \xi \biggl[ -\frac{\mu}{2} {\cal L}^2 - 
\frac{\nu}{8} {\cal L}^2 (f^2 - 1)^2\biggr] + \frac{3}{2} 
\nu \frac{{P'}^2}{\alpha {\cal L}^2}\phi 
\nonumber\\
&+& \frac{3}{2} 
\nu \frac{P'}{\alpha {\cal L}^2} ( \dot{a}_{w} - a_{\theta}') + 
\nu f' ( g' + {g^*}') + \frac{\nu}{4} {\cal L}^2 f ( f^2 -1 ) ( g + g^*),
\label{tww}\\
\kappa \delta \tau_{\theta\theta} &=& 2 \phi \biggl[ 
-\frac{\mu}{2} {\cal L}^2 - \frac{\nu}{8} {\cal L}^2(f^2 - 1)^2\biggr]
+ \frac{3}{2} 
\nu \frac{{P'}^2}{\alpha {\cal L}^2}\xi- 2 \nu a_{\theta}  P f^2
\nonumber\\
&+& \frac{3}{2} 
\nu \frac{P'}{\alpha {\cal L}^2} ( \dot{a}_{w} - a_{\theta}')
- i \nu P f ( \dot{g} - \dot{g}^*) + 
\nu [ P^2 f +\frac{{\cal L}^2}{4} f ( f^2 -1)]( g+ g^*) ,
\label{tthth}\\
\kappa \delta \tau_{\theta w} &=& \biggl[ - \frac{\mu}{2}{\cal L}^2
- \frac{3}{4} \frac{\nu{P'}^2}{\alpha {\cal L}^2} - \frac{\nu}{8} {\cal L}^2
(f^2 -1)^2 \biggr] \pi 
\nonumber\\
&+& \frac{\nu}{2} f' ( \dot{g} + \dot{g}^*) -\frac{i\nu}{2} P [ f(g' -{g^*}')
- f' ( g - g^*) ] - \nu P f^2 a_w.
\label{tthw}
\end{eqnarray}
In Eqs. (\ref{tmn})--(\ref{tthw}) the notations of Eqs. 
(\ref{ff1})--(\ref{ff3}) have been used.
Both Eqs. (\ref{munu})--(\ref{thetwp}) 
and  Eqs. (\ref{tmn})--(\ref{tthw}) are written in general terms 
and no gauge-fixing has been invoked. The gauge-invariant 
equations of motion for the fluctuations are then obtained 
by writing, component by component, Eq. (\ref{einper}) using the 
explicit expressions reported in this Appendix. Then, the gauge-invariant 
fluctuations obtained in Eqs. (\ref{PSI})-(\ref{scgi}), Eqs.
(\ref{V})--(\ref{Z}), Eqs. (\ref{AW})--(\ref{Adiv})
and Eq. (\ref{giD}) 
are inserted in the various components of Eq. (\ref{einper}). The 
final results of this procedure are 
Eqs. (\ref{tens2}) [see Section V], 
Eqs. (\ref{v1})--(\ref{v3}) [see Section VI]  and 
Eqs. (\ref{mn})--(\ref{thth}) [see Section VII].
The same procedure, using the explicit expressions of Eqs. (\ref{deltach}) 
has to be carried on in the case of Eqs. (\ref{phper}) 
and (\ref{Aper}), namely the 
evolution equations for the Higgs and gauge field fluctuations whose 
explicit expressions are also reported in the bulk of the paper.

\renewcommand{\theequation}{B.\arabic{equation}}
\setcounter{equation}{0}
\section{Localization of the vector modes in the polar coordinate system}

In this Appendix the evolution of the vector modes of the geometry 
will be studied in the coordinate system defined by the line element
\begin{equation}
ds^2 = M^2(\rho) [ dt^2 - d\vec{x}^2 ] - d\rho^2 - L^2(\rho) d\theta^2.
\label{cs3}
\end{equation}
Moreover, in order to check for the consistency of our result, it will also be 
done in a specific gauge, namely the gauge where $\zeta_{\mu} = f_{\mu}$.

Consider the perturbed line element for vector fluctuations in the form
 \footnote{ 
In order not to make the formulas too heavy {\em only in this Appendix} 
we denoted by the prime the derivation with respect to 
$x = \sqrt{\lambda} v \rho$. On the other hand, in the bulk of the paper 
the prime denotes the derivation with respect to $w$.} 
\begin{equation}
\delta G_{A B}=\left(\matrix{0
& M D_{\mu}  & L M Q_{\mu} \cr
 M  D_{\mu} & 0 & 0&\cr
L M Q_{\mu} & 0 & 0&\cr}\right).
\label{lorfx}
\end{equation}
The perturbed line element of Eq. (\ref{lorfx}) can be obtained by considering 
the pure vector modes of Eq. (\ref{lorf}) in the gauge $f_{\mu} =\zeta_{\mu}$
and by recalling that $d\rho= L(w) dw$.

In this set-up the evolution of the vector modes is given 
by the following equations:
\begin{eqnarray}
&& D_{\mu}' + ( 3 H + F) D_{\mu}  + \frac{\dot{Q}_{\mu}}{{\cal L}} =0,
\label{v1d}\\
&& \frac{\ddot{D}_{\mu}}{{\cal L}^2} - \frac{2}{m_{H}^2 ~M^2}\Box D_{\mu} - 
\frac{\dot{Q}'_{\mu}}{{\cal L}} + ( H - F) \frac{\dot{Q}_{\mu}}{{\cal L}} = 
2 \frac{L}{M} \frac{\nu}{\alpha} \frac{P'}{{\cal L}^2} 
\frac{\dot{a}_{\mu}}{{\cal L}},
\label{v2d}\\
&& Q_{\mu}'' + ( 4 H + F) Q_{\mu}' + ( F' - H' + 5 H F - 5 H^2) Q_{\mu} 
- \frac{ 2}{ m_{H}^2 ~M^2} \Box Q_{\mu}
\nonumber\\
&& - \frac{1}{{\cal L}} \biggl[ \dot{D}_{\mu}' + ( 5 H 
- F) \dot{D}_{\mu}\biggr] + 2 \frac{L}{M} \biggl( \frac{\nu}{\alpha} 
\frac{P'}{{\cal L}^2} a_{\mu}' + \frac{\nu P f^2}{{\cal L}^2} a_{\mu} \biggr) =0,
\label{v3d}\\
&& a_{\mu}'' + \frac{\ddot{a}_{\mu}}{{\cal L}^2} - \frac{2}{m_{H}^2~M^2} \Box
a_{\mu} + (2 H + F) a_{\mu}'
\nonumber\\
&&- P' \frac{M}{L} \biggl[ Q_{\mu}' - (H- F) Q_{\mu} - \frac{\dot{D}_{\mu}}{{\cal L}}
\biggr] - \alpha a_{\mu} f^2 =0.
\label{v4d}
\end{eqnarray}
In Eqs. (\ref{v1d})--(\ref{v4d}) , $H = d \ln{M}/dx$ and $F= d\ln{L}/dx$
and $a_{\mu}$ is the gauge field fluctuation. 
Defining now the following variables
\begin{equation}
u_{\mu} = \varepsilon \frac{\dot{D}_{\mu}}{{\cal L}} - ( \varepsilon Q_{\mu})',
\label{u}
\end{equation}
where $\varepsilon = L/M$ eqs. (\ref{v1d})--(\ref{v4d}) 
can be written in the following form 
\begin{eqnarray}
&& D_{\mu}' + ( 4 H + \frac{\varepsilon'}{\varepsilon}) D_{\mu} + 
\frac{\dot{Q}_{\mu}}{{\cal L}}=0,
\label{v1e}\\
&& \frac{\dot{u}_{\mu}}{{\cal L}} - \frac{2}{m_{H}^2} 
\frac{\varepsilon}{M^2} \Box D_{\mu} = 2 \varepsilon^2 \frac{\nu}{\alpha} 
P' \frac{\dot{a}_{\mu}}{{\cal L}},
\label{v2e}\\
&& u_{\mu}' + ( 5 H - \frac{\varepsilon'}{\varepsilon}) u_{\mu} + 
2 \frac{\varepsilon}{m_{H}^2 M^2} \Box Q_{\mu} = 2 \varepsilon^2 \biggl[ 
\frac{\nu}{\alpha} P' a_{\mu}' + \frac{\nu P f^2}{{\cal L}^2} a_{\mu}\biggr],
\label{v3e}\\
&& a_{\mu}'' + \frac{\ddot{a}_{\mu}}{{\cal L}^2} - \frac{2}{m_{H}^2 ~M^2} 
\Box a_{\mu} + ( 2 H + F) a_{\mu}' + \frac{P'}{\varepsilon^2}u_{\mu} - 
\alpha f^2 a_{\mu}=0.
\label{v4e}
\end{eqnarray}
Consider now the case where the masses of $D_{\mu}$ and $Q_{\mu}$ are both 
vanishing.  Then we have that 
\begin{equation}
u_{\mu} = 2 \varepsilon^2 \frac{\nu}{\alpha}  \frac{P'}{{\cal L}^2} a_{\mu}. 
\end{equation}
Correspondingly the equation for the gauge field fluctuation can be 
simplified
\begin{equation}
b_{\mu}'' + \frac{\ddot{b}_{\mu}}{{\cal L}^2} - \frac{2}{m_{H}^2 M^2} \Box b_{\mu} 
- \frac{ (M \sqrt{L} P)''}{M \sqrt{L} P} b_{\mu} =0,
\end{equation}
where $b_{\mu} = M \sqrt{L} a_{\mu}$. The zero modes of the system are 
\begin{eqnarray}
&& a_{\mu} = k_{1,\mu} P, 
\label{0a}
\nonumber\\
&& D_{\mu} = \frac{c_{1,\mu}}{M^3 L},
\label{0d}\\
&& Q_{\mu} =   k_{1,\mu} \frac{L}{M} + c_{2,\mu} \frac{M}{L}. 
\label{0q}
\end{eqnarray}
By  now perturbing the action to second order the correct canonical 
normalization of the fields can be deduced. The kinetic terms  
of $a_{\mu}$, $D_{\mu}$ and $Q_{\mu}$ appear in the action 
in the following canonical form 
\begin{eqnarray}
&&\overline{a}_{\mu} = \sqrt{L} a_{\mu},
\nonumber\\
&&\overline{D}_{\mu} = M\sqrt{L} D_{\mu},
\nonumber\\
&&\overline{Q}_{\mu} = M \sqrt{L} Q_{\mu}. 
\end{eqnarray}
Hence, the normalization integrals which should 
converge are 
\begin{eqnarray}
&& |k_{1,\mu}|^2 \int_{0}^{\infty} L(x) P^2(x) dx ,
\label{x1}\\
&& |c_{1,\mu}|^2 \int_{0}^{\infty} \frac{ d x}{M^4(x) L(x)},
\label{x2}\\
&& \int_{0}^{\infty} \biggl[   | k_{1,\mu}|^2 L^3(x) +   
|c_{2,\mu}|^2 \frac{M^4(x)}{L(x)} + 2
k_{1,\mu} c_{1,\mu} M^2(x) L(x) \biggr] dx.
\label{x3}
\end{eqnarray}
Since for $x\to 0$ $L(x) \simeq x$, the integrand of Eq. (\ref{x1}) converges 
in the 
core. It converges also at infinity where, in this 
coordinate system, $P(x) \sim 
e^{-\sqrt{\alpha} x}$ and $L(x) \sim e^{ - c x}$. Since for $x \to \infty$ the warp 
factors are exponentially decreasing the integrand of Eq. (\ref{x2}) 
diverges. Finally the second term of the integrand of Eq. (\ref{x3}) 
diverges as $1/x$ for $x\to 0$ leading to an integral which is 
logarithmically divergent in the same limit.

Recalling  that $dx = {\cal L}(w) d w$
(where ${\cal L} = \sqrt{\lambda} v L$)
 and transforming, accordingly, the limits 
of integration, we get the same result obtained in the bulk of the 
paper taking into account that, in our gauge,
${\cal A}_{\mu} = a_{\mu}$ ,
$V_{\mu} = D_{\mu}$ and $Z_{\mu} = Q_{\mu}$. With this observation 
we clearly see that the normalization integrals of Eqs. (\ref{x1})--
(\ref{x3})  become the same as the ones of Eqs. (\ref{norm1})--(\ref{norm2}) 
and (\ref{norm3}). This consistency check shows also that the 
canonical normalizations have been correctly derived in the two 
parametrizations of the background geometry.

\renewcommand{\theequation}{C.\arabic{equation}}
\setcounter{equation}{0}
\section{Longitudinal system}

For the lowest angular momentum eigenstate Eq. (\ref{mw}) allows to
express ${\cal A}_{\theta}$ as a function of the other fluctuations:
\begin{equation}
\Phi' + ({\cal F}-{\cal H}) \Phi  - 
(3{\cal H} + {\cal F}) \Xi - 3 \Psi' =
\nu f' \Delta_1 - \frac{\nu P'}{\alpha {\cal L}^2} 
{\cal A}_\theta.
\label{athred}
\end{equation}
Using Eq. (\ref{mn}), Eq. (\ref{athred}) can be 
expressed as 
\begin{equation}
\frac{\nu}{\alpha} \frac{P'}{{\cal L}^2}{\cal A}_{\theta} = \Xi' + \Psi' + 2 ({\cal H}+ {\cal F}) \Xi 
 \nu f' \Delta_{1}.
\label{Ith}
\end{equation}
Inserting Eq. (\ref{Ith}) into Eqs. (\ref{real}), (\ref{mn2}), (\ref{ww}) and (\ref{thth}) 
we get 
\begin{eqnarray}
&& \Delta_{1}'' + 4 {\cal H}\Delta_{1}' - \biggl[ P^2 + {\cal L}^2 ( 3 f^2 -1) 
- \frac{2 \alpha {\cal L}^2 P f f'}{P'} \biggr] \Delta_{1}
\nonumber\\
&& + 2 \biggl( f' + \frac{\alpha {\cal L}^2 P f}{\nu P'}\biggr) ( \Psi' + \Xi') + 
2 \biggl[ f'' + 4 {\cal H} f' + f P^2 + \frac{ 2 \alpha {\cal L}^2 P f}{\nu P'}({\cal H} + {\cal F} )\biggr] 
\Xi= 0,
\end{eqnarray}
for the real part of the Higgs equation and 
\begin{eqnarray}
&& \Xi''+3 \Psi''  + \biggl( 18 {\cal H} - 2 {\cal F} - \frac{\alpha {\cal L}^2 f^2 P}{P'} \biggr) \Psi' 
+ \biggl( 10 {\cal H} + 2 {\cal F} -  {\cal F} - \frac{\alpha {\cal L}^2 f^2 P}{P'} \biggr) \Xi'+ 
\nonumber\\
&& 2\biggl[ 3 {\cal H}' + {\cal F}' + 4 {\cal H} ( 3 {\cal H} + {\cal F}) - ({\cal H} + {\cal F})  
\frac{\alpha {\cal L}^2 f^2 P}{P'}\biggr] \Xi + \nu f' \Delta_{1}' 
\nonumber\\
&& + \nu \biggl[ f'' + \biggl( 4 {\cal H} -  \frac{\alpha {\cal L}^2 f^2 P}{P'}\biggr) f' 
+ {\cal L}^2 f ( f^2 - 1) \biggr] 
\Delta_1 =0,
\end{eqnarray}
\begin{eqnarray}
&&\Xi'' - \Psi'' + \biggl[ 2 {\cal H} + 6 {\cal F} - 3  \frac{\alpha {\cal L}^2 f^2 P}{P'}\biggr] \Psi' + 
\biggl( 10 {\cal H} + 2 {\cal F} - 3  \frac{\alpha {\cal L}^2 f^2 P}{P'}\biggr) \Xi 
\nonumber\\
&& +  2\biggl[ 3 {\cal H}' + {\cal F}' + 4 {\cal H} ( 3 {\cal H} + {\cal F}) - 2 \nu P^2 
f^2 - 3({\cal H} + {\cal F})  
\frac{\alpha {\cal L}^2 f^2 P}{P'}\biggr] \Xi  - \nu f' \Delta_{1}' 
\nonumber\\
&&+ \nu \biggl[ 3 f'' + 3 \biggl( 4 {\cal H} -  \frac{\alpha {\cal L}^2 f^2 P}{P'}\biggr) f'
-  {\cal L}^2 f (f^2 - 1)\biggr] 
\Delta_1 =0,
\end{eqnarray}
\begin{eqnarray}
&& 7 \Psi'' + \Xi'' + \biggl( 34 {\cal H} - 10 {\cal F} + \frac{\alpha {\cal L}^2 f^2 P}{P'}\biggr) \Psi' + 
\biggl( 10 {\cal H} + 2 {\cal F} + \frac{\alpha {\cal L}^2 f^2 P}{P'}\biggr) \Xi'
\nonumber\\
&& +  2\biggl[ 3 {\cal H}' + {\cal F}' + 4 {\cal H} ( 3 {\cal H} + {\cal F}) + 2 \nu P^2 
f^2 +({\cal H} + {\cal F})  
\frac{\alpha {\cal L}^2 f^2 P}{P'}\biggr] \Xi  + 3 \nu f' \Delta_{1}' 
\nonumber\\
&& - \nu \biggl[ f'' + \biggl( 4 {\cal H} -  \frac{\alpha {\cal L}^2 f^2 P}{P'}\biggr) f' -
3  {\cal L}^2 f (f^2 - 1)\biggr] 
\Delta_1 =0,
\end{eqnarray}
for the remaining components of the perturbed Einstein equations. These equations can be 
used in order to study the asymptotic 
behaviour of the zero mode of $\Psi$ near the core of the vortex. By using the 
asymptotic expressions for $w\to -\infty$ it is found, as expected, that $\Psi \simeq
 {\rm constant}$, 
$\Xi \simeq e^{-2 w}$, and $\Delta_{1} \simeq e^{w}$. 
\end{appendix}


\newpage

\begin{thebibliography}{99}

\bibitem{m1}  V. Rubakov and M. Shaposhnikov, Phys. Lett. B {\bf 125} 
136 (1982).

\bibitem{kap} D. B. Kaplan, Phys. Lett. B {\bf 288}, 342 (1992).

\bibitem{nin} H. B. Nielsen and M. Ninomya, Phys. Lett. 
B {\bf 105}, 219 (1985). 

\bibitem{wg} P. H. Ginsparg, and K. G. Wilson, Phys. Rev. D {\bf 25}, 
2649 (1982).

\bibitem{tro1} M.V. Libanov and 
S.V. Troitsky Nucl. Phys. B {\bf 599} 319 (2001); J. M. Fr\`ere, 
 M.V. Libanov and S.V. Troitsky, Phys. Lett. B {\bf 512}, 169 (2001). 

\bibitem{tro2} J.M. Fr\`ere, 
M.V. Libanov, S.V. Troitsky, JHEP {\bf 0111} 025, (2001). 

\bibitem{m2} V. Rubakov and M. Shaposhnikov, Phys. Lett. B {\bf 125}, 139 
(1983).

\bibitem{ak} K. Akama, in {\em Proceedings of the 
Symposium on Gauge Theory and Gravitation}, Nara, Japan, 
eds. K. Kikkawa, N. Nakanishi and H. Nariai, (Springer-Verlag, 1983),
[hep-th/0001113].

\bibitem{vis} M. Visser, Phys. Lett. B {\bf 159}, 22 (1985).

\bibitem{ran} S. Randjbar-Daemi and 
C. Wetterich, Phys. Lett. B {\bf 166}, 65 (1986).

\bibitem{rs1} L. Randall and R. Sundrum, Phys. 
Rev. Lett. {\bf 83} 3370 (1999). 

\bibitem{rs2}  L. Randall and R. Sundrum, Phys. 
Rev. Lett. {\bf 83} 4690 (1999). 
 
\bibitem{kt1} A. Kehagias and K. Tamvakis, Phys.Lett. B {\bf 504}, 38 (2001).

\bibitem{kt2} A. Kehagias and K. Tamvakis, hep-th/0011006.

\bibitem{gremm1} M. Gremm, Phys. Lett. B {\bf 478}, 434 (2000).

\bibitem{gremm2} M. Gremm,  Phys. Rev. D {\bf 62}, 044017 (2000).

\bibitem{free} O. DeWolfe, D.Z. Freedman, S.S. Gubser, A. Karch, 
Phys.Rev.D {\bf 62}, 046008 (2000).  

\bibitem{free2} O. DeWolfe, D. Z. Freedman, hep-th/0002226. 

\bibitem{mg1} M. Giovannini, Phys. Rev. D {\bf 64}, 064023 (2001); 
Phys.Rev. D {\bf 64}, 124004 (2001).

\bibitem{mg2} M. Giovannini, Phys. Rev. D {\bf 65}, 064008 (2002).

\bibitem{def} A. G. Cohen and D. B. Kaplan, Phys. Lett. B {\bf 470}, 52 
(1999).

\bibitem{def1}A. Chodos and E. Poppitz, Phys. Lett. B {\bf 471}, 119 (1999). 

\bibitem{def2}I. Olasagasti and A. Vilenkin, Phys. Rev. D {\bf 62}, 044014
(2000).

\bibitem{def3}  R. Gregory, Phys. Rev. Lett. {\bf 84}, 2564 (2000).

\bibitem{def3b} C. Burgess, J.  Cline, N. R. Constable, and  
H. Firouzjahi, JHEP {\bf 0201}, 014 (2002). 

\bibitem{gs} T. Gherghetta and M. Shaposhnikov, Phys.Rev.Lett. {\bf 85}, 240 
(2000). 

\bibitem{grs} T. Gherghetta, E. 
Roessl, M. Shaposhnikov, Phys.Lett.B {\bf 491}, 353 (2000). 

\bibitem{gm} M. Giovannini, Phys. Rev. D {\bf 63}, 085005 (2001);  
Phys. Rev. D {\bf 63}, 064011 (2001). 

\bibitem{Dvali:2000ty}
G.~Dvali,
hep-th/0004057.
\bibitem{Randjbar-Daemi:1983qa}
S.~Randjbar-Daemi, A.~Salam and J.~Strathdee,
Phys.\ Lett.\ B {\bf 132}, 56 (1983).

\bibitem{Randjbar-Daemi:2000cr}
S.~Randjbar-Daemi and M.~Shaposhnikov,
Phys.\ Lett.\ B {\bf 492}, 361 (2000).

\bibitem{Randjbar-Daemi:2000ft}
S.~Randjbar-Daemi and M.~Shaposhnikov,
Phys.\ Lett.\ B {\bf 491}, 329 (2000).

\bibitem{def3a} R. 
Gregory and A. Padilla,  Class. Quant. Grav. {\bf 19}, 279 (2002).

\bibitem{mhm1} M. Giovannini, H. Meyer, M. E. 
Shaposhnikov, Nucl.Phys. B {\bf 619}, 615 (2001)

\bibitem{mhm2} M. Giovannini, H.B. Meyer, Phys.Rev.D {\bf 64}, 124025 (2001). 

\bibitem{no} H.B.~Nielsen and P.~Olesen, Nucl. Phys.B {\bf 61}, 45 (1973).  

\bibitem{bar} J. M. Bardeen, Phys. Rev. D {\bf 22}, 1882 (1980).

\bibitem{ner} A. Neronov, Phys.Rev. D {\bf 64}, 044019 (2001). 

\bibitem{mgv1} M. Giovannini, hep-th/0204235 (to appear in Phys. Rev. D).

\bibitem{susyqm} F. Cooper, A. Khare, and U. Sukhatme, Phys. Rep. {\bf 251}, 267 (1995).

\bibitem{lib} M. V. Libanov and  E. Y. Nougaev,  hep-ph/0201162.


\end{thebibliography}
\end{document}